\documentclass[twocolumn,showpacs]{revtex4}

\usepackage{graphicx}

\begin{document}

\title{Pulsed squeezed-light generation in a waveguide with
second-subharmonic generation and periodic corrugation}

\author{Jan Pe\v{r}ina Jr.}
\address{RCPTM, Joint Laboratory of Optics of Palack\'{y} University
and Institute of Physics of Academy of Sciences of the Czech
Republic, 17. listopadu 12, 771~46 Olomouc, Czech Republic}

\email{perinaj@prfnw.upol.cz}

\begin{abstract}
Quantum pulsed second-subharmonic generation in a planar waveguide
with a small periodic corrugation at the surface is studied.
Back-scattering of the interacting fields on the corrugation
enhances the nonlinear interaction giving larger values of
squeezing. The problem of back-scattering is treated by
perturbation theory, using the Fourier transform for
non-dispersion propagation, and by numerical approach in the
general case. Optimum spectral modes for squeezed-light generation
are found using the Bloch-Messiah reduction. Improvement in
squeezing and increase of numbers of generated photons are
quantified for the corrugation resonating with the fundamental and
second-subharmonic field. Splitting of the generated pulse by the
corrugation is predicted.
\end{abstract}



\maketitle

\section{Introduction}

The process of second-subharmonic generation which is just the
inverse process to that of second-harmonic generation
\cite{Armstrong1962} is interesting not only as a means of
frequency conversion. It can also serve as an efficient source of
squeezed light \cite{Luks1988} which amplitude fluctuations are
suppressed below the limit given by quantum uncertainty relations
(for a review, see, e.g.,
\cite{Perina1991,Mandel1995,Dodonov2002,Vogel1994}). Such
nonclassical light can be emitted both in the fundamental as well
as second-subharmonic frequency (SSF) fields \cite{Bachor2004}.
Also light with nonclassical photon-number statistics can be
obtained in this process under suitable conditions
\cite{Schleich1987,Perina1990}. The process of second-subharmonic
generation belongs to the whole family of optical parametric
processes which share many common features
\cite{Butcher1991,Boyd1994,Mitchell2009}. Among them, spontaneous
parametric down-conversion with its ability to generate entangled
photon pairs plays an important role \cite{Mandel1995}.

In homogeneous nonlinear media, the perfectly phase matched
nonlinear interaction gives the largest possible squeezing of
amplitude fluctuations. The principal squeeze variance of the SSF
field asymptotically reaches zero for large values of the gain of
nonlinear interaction \cite{Ou1994}. That is why enhancement of
the effective nonlinearity is important as it immediately results
in larger squeezing and lower pumping intensities needed for
reaching a given value of squeezing. For this purpose,
configurations with cavities filled by a nonlinear medium have
been usually used to generate squeezed light (e.g.,
\cite{Bachor2004,Lawrence2002}). Also nonlinear waveguides, for
which strong spatial localization of optical fields in the
transverse plane is characteristic, profit from enhancement of the
effective nonlinearity. This enhancement has been widely exploited
when generating photon pairs in classical
\cite{PerinaJr2008,Mosley2009,Karpinski2009}, multi-layer
\cite{Lanco2006}, Bragg-reflection \cite{Svozilik2011,Horn2012}
and photonic-wire \cite{Matsuda2012} waveguides. Effective
nonlinearity in a waveguide can also be increased by the use of
mode coupling through evanescent waves with fields in a
neighboring waveguide. This occurs due to an additional spatial
modulation induced by energy exchange
\cite{Dong2004,PerinaJr2000}. Additional spatial modulation can
also be introduced in a simpler geometry using a linear periodic
corrugation at or below the waveguide surface
\cite{Joannopoulos1995,Bertolotti2001}. Back-scattering occurring
at the corrugation modifies electric-field amplitudes of the
nonlinearly interacting fields which results in the enhancement of
effective nonlinearity under suitable conditions
\cite{Haus1998,Pezzetta2001,PerinaJr2007a}. Back-scattering on a
periodic corrugation has already been exploited for cw
squeezed-light generation both in the process of second-harmonic
\cite{Tricca2004} and second-subharmonic generation
\cite{PerinaJr2007a}. Scattering on a periodic corrugation in a
waveguide can also be used to enhance second-harmonic generation
in \v{C}erenkov configuration
\cite{Ctyroky2000,Kotacka2001,Pezzetta2002}. Squeezed-light
generation in nonlinear photonic structures has been discussed in
general in \cite{Sakoda2002}.

Photonic structures modify in general phase matching conditions of
the nonlinear interaction. Propagation constants in waveguiding
structures represent a typical example. Especially
Bragg-reflection waveguides offer wide possibilities in this
direction \cite{Svozilik2012,Kang2012}. Also scattering on a
periodic corrugation gives an additional term to nonlinear
phase-matching conditions
\cite{PerinaJr2004,PerinaJr2005,PerinaJr2007}. For this reason we
need a tool that allows us to reach nonlinear phase matching
conditions for an arbitrary photonic structure. Periodic poling
\cite{Serkland1995,Serkland1997,Yu2005,Huang2006} of $ \chi^{(2)}
$ susceptibility has occurred to be extraordinarily useful here
and has resulted in the so-called quasi-phase-matched nonlinear
interactions. Using this method even spectrally broad-band
two-mode nonlinear interaction with femtosecond pulses has become
possible \cite{Schober2005}.

Due to temporal energy concentration, the pulsed regime of the
nonlinear process allows to use lower pumping powers to observe
the needed level of squeezing. It also brings into attention new
features of the generated light, namely its spectral modal
structure. In the pulsed regime and travelling-wave configuration,
squeezed-light generation in the considered nonlinear interaction
has been studied with the help of phase-space quasi-distributions
or the corresponding Langevin stochastic equations
\cite{Werner1995}. A local oscillator in the form of an ultrashort
optical pulse is needed to observe pulsed squeezing experimentally
using homodyne detection. The effort to observe the largest
possible values of squeezing has raised the question about an
optimum shape of the local-oscillator field
\cite{Kumar1990,Shapiro1997,Bennink2002}. The Bloch-Messiah
reduction of the evolution matrices (operators) has been proposed
for the solution of the modal structure \cite{Braunstein2005}.
This method has been elaborated in detail in
\cite{Wasilewski2006,Lvovsky2007} for degenerate parametric
down-conversion in a BBO crystal. Also the relation between the
Bloch-Messiah reduction and Schmidt decomposition of a two-photon
spectral amplitude characterizing spontaneous parametric
down-conversion has been found. The results obtained in
single-pass geometry have been generalized to the nonlinear
interaction in a cavity \cite{Patera2010}. Mode structure of
nonclassical states arising from squeezed states after
post-selection done with on/off detectors has been analyzed in
\cite{Sasaki2006}. The discussed effects occurring in temporal (or
spectral) domains have their counterparts in spatial domain, in
the transverse plane of the beams. Also here eigenmodes typical
for the nonlinear interaction can be revealed
\cite{Annamalai2011}. Circular symmetry of the usual optical
beams, however, results in different types of eigenmodes defined
in the transverse plane. Despite this, a close similarity in the
behavior of fields in the spectral and spatial domains can be
found. As an important example, correlations between intensities
of the interacting fields in their transverse planes can be
mentioned \cite{Scotto2003,Bache2002}.

Here, continuing the investigation in \cite{PerinaJr2007a} we pay
attention to the pulsed SSF generation in a waveguide with linear
periodic corrugation on its surface that causes back-scattering of
the interacting fields. Assuming strong pulsed fundamental field,
we study squeezed-light generation in the SSF field. We pay our
attention to the enhancement of nonlinearity and the related
increase of squeezing applying three different approaches. We
utilize a sophisticated perturbation approach, the method of
Fourier transformation in non-dispersion field propagation, and
numerical solution for the general case. Squeezing is
characterized by a principal squeeze variance introduced in
\cite{Luks1988,Perina1991,Dodonov2002} that is determined for
suitable spectral modes.

The paper is organized as follows. In Sec.~II, a multi-mode
quantum model of the nonlinear interaction using the appropriate
momentum operator and the related Heisenberg equations is
presented. Sec.~III is devoted to the perturbation solution of the
model that is divided into three parts. A general perturbation
solution is found in Subsec.~IIIA, Gaussian spectral approximation
to the solution is applied in Subsec.~IIIB, and squeezing is
analyzed in Subsec.~IIIC. If inter-mode dispersion is omitted, the
model can be solved by the Fourier transform, as shown in Sec.~IV.
Discrete formulation of the model is elaborated in Sec.~V devoted
to the numerical solution. Also quantities useful in the
characterization of the generated field are introduced in this
section. Discussion of squeezing and appropriate spectral
eigenmodes is contained in Sec.~VI. Whereas the model with
non-dispersion field propagation is analyzed in Subsec.~VIA, the
results of the general approach are studied in Subsec.~VIB.
Conclusions are drawn in Sec.~VII. An optimum mode profile giving
the maximum pulsed squeezing is found in Appendix A.

\section{Quantum multi-mode model of second-subharmonic generation}

\begin{figure}    
 \resizebox{0.9\hsize}{!}{\includegraphics{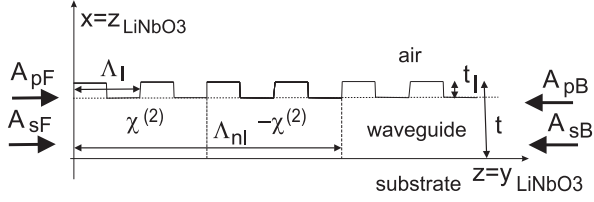}}

 \vspace{2mm}
 \caption{Scheme of a periodically-poled nonlinear waveguide with
 a rectangular transverse profile (thickness $ t $ and width $ \Delta y
 $) and length $ L $ having $ \chi^{(2)} $ susceptibility. A linear corrugation
 with period $ \Lambda_l $ and depth $ t_l $ occurs on the waveguide upper
 surface; $ \Lambda_{nl} $ denotes period of nonlinear poling.
 The waveguide is made of LiNbO$ {}_3 $ which optical axis coincides with the $ x
 $ axis. Four fields interact inside the waveguide: forward-propagating fundamental
 (electric-field amplitude $ A_{p_F} $), backward-propagating fundamental
 ($ A_{p_B} $), forward-propagating second-subharmonic ($ A_{s_F} $),
 and backward-propagating second-subharmonic ($ A_{s_B} $) fields.}
 \label{fig1}
\end{figure}

We consider a nonlinear waveguide made of LiNbO$ {}_3 $ (see
Fig.~\ref{fig1}) in the configuration that allows to generate a
second-subharmonic (SSF) field using $ \chi^{(2)} $ nonlinearity
and pumping, e.g., by the second-harmonic of a Nd:YAG laser at the
wavelength of $ \lambda_s = 1.064 \times 10^{-6} $~m. Under a
suitable choice of waveguide parameters, the waveguide is
single-mode for both the fundamental and SSF field and allows for
efficient nonlinear interaction. A linear corrugation fabricated
at the top of the waveguide leads to back-scattering of the
interacting fields which results in the enhancement of
electric-field amplitudes inside the waveguide under suitable
conditions. This leads to effective increase of the nonlinear
interaction and gives larger amount of squeezing of the SSF light.
Quasi-phase matching of the nonlinear interaction is guaranteed by
periodic poling with an appropriate poling period. A detailed
description of the waveguide was given in \cite{PerinaJr2007a}. It
has been shown that the investigated waveguide can be described by
the following parameters: propagation and coupling constants of
the fundamental and SSF fields and constants characterizing the
nonlinear interaction occurring among both the forward- and
backward-propagating fields. When losses inside the waveguide
caused both by absorption and scattering of the light outside the
guided modes are neglected we can describe the nonlinear
interaction in the waveguide as follows.

According to quantum theory, the electric-field vector operator
amplitudes $ \hat{\bf E}_a(x,y,z,t) $ ($ a=p,s $) at time $ t $
and spatial point $ (x,y,z) $ inside the waveguide can be
decomposed into harmonic plane waves with mode operator amplitudes
$ \hat{a}_a $ in the Heisenberg picture
\cite{Vogel1994,Perina1990}:
\begin{eqnarray}    
 \hat{\bf E}_a(x,y,z,t) &=& \int_{0}^{\infty} d\omega_a
  \hat{\bf E}_a(x,y,z,\omega_a) \exp(-i\omega_a t) , \nonumber \\
 & &
\label{1} \\
 \hat{\bf E}_a(x,y,z,\omega_a) &=& i\left[ \hat{a}_{a_F}(z,\omega_a)
  {\bf e}_a(x,y,\omega_a) \right. \nonumber \\
  & & \hspace{-20mm} \mbox{} + \left. \hat{a}_{a_B}(z,\omega_a)
  {\bf e}_a(x,y,\omega_a) - {\rm H.c.} \right] ,\hspace{2mm}
  a=p,s.
\label{2}
\end{eqnarray}
In Eq.~(\ref{2}), $ \hat{a}_{a_F}(z,\omega_a ) $ [$
\hat{a}_{a_B}(z,\omega_a ) $] denotes an annihilation operator of
the mode with frequency $ \omega_a $ in field $ a $ propagating
forward [backward]. We note that the mode vector functions $ {\bf
e}_a $ as well as the corresponding propagation constants $
\beta_a $ depend on the frequency $ \omega_a $ and their form can
be found in \cite{PerinaJr2007a}.

Evolution of the nonlinearly-interacting quantum optical fields
inside the waveguide is described by the following momentum
operator $ \hat{G} $ \cite{Perina1990,PerinaJr2000}:
\begin{eqnarray}     
 \hat{G}(z) &=& \nonumber \\
 & & \hspace{-12mm} \mbox{}
  \sum_{a=p,s} \sum_{b=F,B} \int_{0}^{\infty} d\omega_a
  (\pm 1)_b \hbar \beta_{a_b}(\omega_a) \hat{a}^\dagger_{a_b}(z,\omega_a)
  \hat{a}_{a_b}(z,\omega_a)\nonumber \\
 & & \hspace{-12mm} \mbox{}
  + \Bigl[ \sum_{a=p,s} \int_{0}^{\infty} d\omega_a \hbar
  K_a(\omega_a)\exp\left(i\frac{2\pi}{\Lambda_a}z\right) \nonumber \\
 & & \mbox{} \times \hat{a}^\dagger_{a_F}(z,\omega_a)
  \hat{a}_{a_B}(z,\omega_a) + {\rm H.c.} \Bigr] \nonumber \\
 & & \hspace{-12mm} \mbox{} - \Bigl[ \sum_{b=F,B} 2i \int_{0}^{\infty}
  d\omega_s \int_{0}^{\infty}
  d\omega_s^{'} (\pm 1)_b  K_{nl,q}(\omega_s,\omega_s^{'})
  \nonumber \\
 & & \hspace{-8mm} \mbox{} \times \exp\left(i\frac{2\pi q}{\Lambda_{nl}}z\right)
  \hat{a}^\dagger_{s_b}(z,\omega_s)
  \hat{a}^\dagger_{s_b}(z,\omega_s^{'}) \hat{a}_{p_b}(z,\omega_s+\omega_s^{'})
  \nonumber \\
 & &  \mbox{} + {\rm H.c.} \Bigr] ;
\label{3}
\end{eqnarray}
$ (\pm 1)_F \equiv 1$ and $ (\pm 1)_B \equiv -1 $. The linear
coupling constants $ K_p(\omega_p) $ [$ K_s(\omega_s) $] describe
scattering on the corrugation with period $ \Lambda_p $ [$
\Lambda_s $] that leads to the coupling between the fundamental
[SSF] fields propagating forward and backward. Under suitable
conditions, this scattering leads to increased values of
electric-field amplitudes inside the waveguide. However, this
increase is much lower than that found in layered photonic
band-gap structures
\cite{Scalora1997,DAguanno2001,Dumeige2001,PerinaJr2011}. On the
other hand, the nonlinear coupling constants $
K_{nl,q}(\omega_s,\omega'_s) $ characterize the nonlinear
interaction among co-propagating fields and obey the symmetry
relation $ K_{nl,q}(\omega_s,\omega_s^{'}) =
K_{nl,q}(\omega_s^{'},\omega_s) $. Constant $ \Lambda_{nl} $ gives
the poling period whereas the integer number $ q $ determines a
harmonic frequency employed in the quasi-phase-matching. Planck
constant is denoted as $ \hbar $ and symbol $ {\rm H.c.} $
substitutes the Hermitian conjugated term.

The Heisenberg equations, $ d\hat{X}/dz = - i/\hbar [ \hat{G},
\hat{X} ] $ for an arbitrary operator $ \hat{X} $, written for the
momentum operator $ \hat{G} $ in Eq.~(\ref{3}) attain the form:
\begin{eqnarray}    
 \frac{d\hat{a}_{s_F}(z,\omega_s)}{dz} &=& i\beta_s(\omega_s)\hat{a}_{s_F}(z,\omega_s)
  \nonumber \\
  & & \mbox{} +
  iK_{s}(\omega_s)\exp\left(i\frac{2\pi}{\Lambda_s}z\right) \hat{a}_{s_B}(z,\omega_s)
  \nonumber \\
  & &  \mbox{} + 4\int_{0}^{\infty}d\omega_s^{'}
  K_{nl,q}(\omega_s,\omega_s^{'})\exp\left(i\frac{2\pi q}{\Lambda_{nl}}z\right)
  \nonumber \\
  & & \mbox{} \hspace{10mm} \times \hat{a}_{p_F}(z,\omega_s+\omega_s^{'})
  \hat{a}^\dagger_{s_F}(z,\omega_s^{'}) , \nonumber\\
 \frac{d\hat{a}_{s_B}(z,\omega_s)}{dz} &=& -i\beta_s(\omega_s)\hat{a}_{s_B}(z,\omega_s)
  \nonumber \\
  & & \mbox{} - iK_{s}^*(\omega_s)\exp\left(-i\frac{2\pi}{\Lambda_s}z\right)
   \hat{a}_{s_F}(z,\omega_s) \nonumber \\
  & & \mbox{} - 4\int_{0}^{\infty}d\omega_s^{'}
  K_{nl,q}(\omega_s,\omega_s^{'}) \exp\left(i\frac{2\pi q}{\Lambda_{nl}}z\right)
  \nonumber \\
  & & \mbox{} \hspace{10mm} \times \hat{a}_{p_B}(z,\omega_s+\omega_s^{'})
  \hat{a}^\dagger_{s_B}(z,\omega_s^{'}) , \nonumber\\
 \frac{d\hat{a}_{p_F}(z,\omega_p)}{dz} &=& i\beta_p(\omega_p)\hat{a}_{p_F}(z,\omega_p)
  \nonumber \\
  & & \hspace{-12mm} \mbox{} +
  iK_{p}(\omega_p)\exp\left(i\frac{2\pi}{\Lambda_p}z\right) \hat{a}_{p_B}(z,\omega_p)
  \nonumber \\
  & & \hspace{-12mm} \mbox{} - 2\int_{0}^{\infty}d\omega_s
  K_{nl,q}^*(\omega_s,\omega_p-\omega_s) \exp\left(-i\frac{2\pi q}{\Lambda_{nl}}z\right)
  \nonumber \\
  & & \mbox{} \hspace{-2mm} \times \hat{a}_{s_F}(z,\omega_s)
  \hat{a}_{s_F}(z,\omega_p-\omega_s) , \nonumber\\
 \frac{d\hat{a}_{p_B}(z,\omega_p)}{dz} &=& -i\beta_p(\omega_p)\hat{a}_{p_B}(z,\omega_p)
  \nonumber \\
  & & \hspace{-12mm} \mbox{} - iK_{p}^*(\omega_p)\exp\left(-i\frac{2\pi}{\Lambda_p}z\right)
  \hat{a}_{p_F}(z,\omega_p) \nonumber \\
  & & \hspace{-12mm} \mbox{} + 2\int_{0}^{\infty}d\omega_s
  K_{nl,q}^*(\omega_s,\omega_p-\omega_s) \exp\left(-i\frac{2\pi q}{\Lambda_{nl}}z\right)
  \nonumber \\
  & & \mbox{} \hspace{-2mm} \times \hat{a}_{s_B}(z,\omega_s)
 \hat{a}_{s_B}(z,\omega_p-\omega_s) .
\label{4}
\end{eqnarray}

Creation and annihilation operators of the incident fields are
assumed to fulfil the boson commutation relations, i.e.
\begin{eqnarray}   
 { [\hat{a}_{a_F}(0,\omega_a),
 \hat{a}^\dagger_{a'_F}(0,\omega_{a'})] } &=&
 \delta_{a,a'}\delta(\omega_a -\omega_{a'}) , \nonumber \\
 { [\hat{a}_{a_B}(L,\omega_a),
  \hat{a}^\dagger_{a'_B}(L,\omega_{a'})] } &=&
  \delta_{a,a'}\delta(\omega_a -\omega_{a'}),
  \nonumber \\
  & & \hspace{5mm}  a=p,s .
\label{5}
\end{eqnarray}
The remaining commutators are zero. It has been shown in
\cite{Luis1996} for quadratic momentum operators $ \hat{G} $ that
also the output operators obey the boson commutation relations:
\begin{eqnarray}   
 [\hat{a}_{a_F}(L,\omega_a),
 \hat{a}^\dagger_{a'_F}(L,\omega_{a'})] &=&
 \delta_{a,a'}\delta(\omega_a -\omega_{a'}) , \nonumber \\
 { [\hat{a}_{a_B}(0,\omega_a),
 \hat{a}^\dagger_{a'_B}(0,\omega_{a'})] } &=&
 \delta_{a,a'}\delta(\omega_a -\omega_{a'}), \nonumber \\
 & & \hspace{5mm} a=p,s
\label{6}
\end{eqnarray}
and commutators not mentioned in Eq.~(\ref{6}) are zero.

We note that the nonlinear operator equations written in
Eq.~(\ref{4}) have one integral of motion arising from the
conservation of energy flux:
\begin{eqnarray}     
 & & \frac{d}{dz} \left[ \int_{0}^{\infty} d\omega_s\,
  \hat{a}^\dagger_{s_F}(z,\omega_s)\hat{a}_{s_F}(z,\omega_s)
  \right. \nonumber \\
 & & \hspace{5mm} \mbox{} - \int_{0}^{\infty} d\omega_s\,
  \hat{a}^\dagger_{s_B}(z,\omega_s)\hat{a}_{s_B}(z,\omega_s)
  \nonumber \\
 & & \hspace{5mm} \mbox{}+ 2 \int_{0}^{\infty} d\omega_p\,
   \hat{a}^\dagger_{p_F}(z,\omega_p)\hat{a}_{p_F}(z,\omega_p)
   \nonumber \\
 & & \hspace{5mm} \mbox{}
   - \left. 2 \int_{0}^{\infty} d\omega_p\,
   \hat{a}^\dagger_{p_B}(z,\omega_p)\hat{a}_{p_B}(z,\omega_p)
   \right]  = 0 .
\label{7}
\end{eqnarray}

It is convenient to introduce new operators $ \hat{A} $ that take
into account the harmonic spatial evolution induced by the
corrugation present in general in both the fundamental and SSF
fields, $ \hat{a}_{a_F}(z,\omega_a) = \hat{A}_{a_F}(z,\omega_a)
\exp( i\pi z/\Lambda_a) $ and $ \hat{a}_{a_B}(z,\omega_a) =
\hat{A}_{a_B}(z,\omega_a) \exp(-i \pi z/\Lambda_a) $ for $ a=p,s
$. Equations (\ref{4}) written for the operators $ \hat{A} $ take
the form:
\begin{eqnarray}    
 \frac{d\hat{A}_{s_F}(z,\omega_s)}{dz} &=&
  i\frac{\delta_{s}(\omega_s)}{2} \hat{A}_{s_F}(z,\omega_s) + iK_s(\omega_s)
  \hat{A}_{s_B}(z,\omega_s) \nonumber \\
  & & \hspace{-15mm} \mbox{} + 4\int_{0}^{\infty}d\omega_s^{'} K_{nl,q}(\omega_s,\omega_s^{'})
  \exp[i\delta_{nl,q} z] \nonumber \\
  & & \hspace{-5mm} \mbox{} \times
  \hat{A}_{p_F}(z,\omega_s+\omega_s^{'}) \hat{A}^\dagger_{s_F}(z,\omega_s^{'}) ,
  \nonumber\\
 \frac{d\hat{A}_{s_B}(z,\omega_s)}{dz} &=&
  - i\frac{\delta_{s}(\omega_s)}{2} \hat{A}_{s_B}(z,\omega_s)
  - iK_{s}^*(\omega_s) \hat{A}_{s_F}(z,\omega_s)
  \nonumber \\
  & & \hspace{-15mm} \mbox{} - 4\int_{0}^{\infty}d\omega_s^{'} K_{nl,q}(\omega_s,\omega_s^{'})
  \exp[-i\delta_{nl,q} z] \nonumber \\
  & & \hspace{-5mm} \mbox{} \times
  \hat{A}_{p_B}(z,\omega_s+\omega_s^{'}) \hat{A}^\dagger_{s_B}(z,\omega_s^{'}) ,
\label{8} \\
 \frac{d\hat{A}_{p_F}(z,\omega_p)}{dz} &=&
  i\frac{\delta_{p}(\omega_p)}{2} \hat{A}_{p_F}(z,\omega_p) +
  iK_{p}(\omega_p) \hat{A}_{p_B}(z,\omega_p)
  \nonumber \\
  & & \hspace{-15mm} \mbox{} - 2\int_{0}^{\infty}d\omega_s K_{nl,q}^*(\omega_s,\omega_p-\omega_s)
  \exp[-i\delta_{nl,q} z] \nonumber \\
  & & \hspace{-5mm} \mbox{} \times
   \hat{A}_{s_F}(z,\omega_s) \hat{A}_{s_F}(z,\omega_p-\omega_s) ,
  \nonumber\\
 \frac{d\hat{A}_{p_B}(z,\omega_p)}{dz} &=&
  - i\frac{\delta_{p}(\omega_p)}{2} \hat{A}_{p_B}(z,\omega_p)
  - iK_{p}^*(\omega_p) \hat{A}_{p_F}(z,\omega_p)
  \nonumber \\
  & & \hspace{-15mm} \mbox{} + 2\int_{0}^{\infty}d\omega_s K_{nl,q}^*(\omega_s,\omega_p-\omega_s)
  \exp[i\delta_{nl,q} z] \nonumber \\
  & & \hspace{-5mm} \mbox{} \times
   \hat{A}_{s_B}(z,\omega_s) \hat{A}_{s_B}(z,\omega_p-\omega_s) .
\label{9}
\end{eqnarray}
Linear phase mismatches $ \delta_p $, $ \delta_s $ and nonlinear
phase mismatch $ \delta_{nl,q} $ ($ q=0,\pm 1 $) are defined as:
\begin{eqnarray}   
 \delta_a(\omega_a) &=& 2\beta_a(\omega_a) - \frac{2\pi}{\Lambda_a}, \hspace{3mm} a=p,s
 , \nonumber \\
 \delta_{nl,q} &=& \frac{\pi}{\Lambda_p} - 2 \frac{\pi}{\Lambda_s} +
  \frac{2\pi q}{\Lambda_{nl}} .
\label{10}
\end{eqnarray}
We note that whereas $ q $ was used for the nonlinear interaction
among the forward-propagating fields, $ -q $ was chosen for the
interaction among the backward-propagating fields in
Eqs.~(\ref{8}) and (\ref{9}). We also note that equations similar
to (\ref{8}) and (\ref{9}) can be derived from the wave equation
considering classical fields and invoking paraxial approximation
\cite{Mitchell2009}.

We assume that the fundamental field is strong and its depletion
due to the interaction with the SSF field can be neglected. In
this case, the equations in (\ref{9}) become linear and break into
groups with two linear equations for a given frequency $ \omega_p
$:
\begin{eqnarray}    
 \frac{d\hat{A}_{p_F}(z,\omega_p)}{dz} &=&
  i\frac{\delta_{p}(\omega_p)}{2} \hat{A}_{p_F}(z,\omega_p) +
  iK_{p}(\omega_p) \hat{A}_{p_B}(z,\omega_p),
  \nonumber \\
 \frac{d\hat{A}_{p_B}(z,\omega_p)}{dz} &=&
  - i\frac{\delta_{p}(\omega_p)}{2} \hat{A}_{p_B}(z,\omega_p)
  - iK_{p}^*(\omega_p) \hat{A}_{p_F}(z,\omega_p) . \nonumber \\
 & &
\label{11}
\end{eqnarray}

The solution of Eqs.~(\ref{11}) can be easily obtained in the
matrix form
\begin{eqnarray} 
 \left[ \begin{array}{c} \hat{A}_{p_F}(z,\omega_p) \cr
  \hat{A}_{p_B}(z,\omega_p) \end{array} \right] &=&
  \sum_{\pm} B_p^{\pm}(\omega_p) \exp[\pm i\Delta_p(\omega_p)z]
  \nonumber \\
 & & \mbox{} \times \left[ \begin{array}{c} \hat{A}_{p_F}(0,\omega_p) \cr
  \hat{A}_{p_B}(L,\omega_p) \end{array} \right]
\label{12}
\end{eqnarray}
using the eigenfrequencies $ \pm \Delta_p $, $ \Delta_p(\omega_p)
= \sqrt{\delta_p^2(\omega_p)/4-|K_p(\omega_p)|^2} $. Symbol $
\sum_{\pm}$ occurring in Eq.~(\ref{12}) means the summation over
the terms differing in their signs. We note that the boundary
condition for the backward-propagating field is chosen at $ z=L $.
The matrices $ B_p^{\pm}(\omega_p) $ are defined as
\begin{equation} 
 B_p^{\pm}(\omega_p) = D_p
  \left[ \begin{array}{cc} \left(\pm\frac{\delta_p}{2}+\Delta_p\right)
  \exp(\mp i\Delta_p L) & \pm K_p \cr
  \mp K_p^* \exp(\mp i\Delta_p L)  & \mp \frac{\delta_p}{2}+\Delta_p
  \end{array} \right]
\label{13}
\end{equation}
and $ D_p(\omega_p) = [2\Delta_p\cos(\Delta_p L) -i\delta_p
\sin(\Delta_p L)]^{-1} $.

If the corrugation is missing, the matrices $ B_p^{\pm} $ in
Eq.~(\ref{13}) take a simple form obtainable in the limit $ K_p
\longrightarrow 0 $ together with $ \Lambda_p \longrightarrow
\infty $:
\begin{equation} 
 B_p^{+} = \left[ \begin{array}{cc} 1 & 0 \cr 0 & 0 \end{array}
  \right], \hspace{5mm}
 B_p^{-} = \left[ \begin{array}{cc} 0 & 0 \cr 0 & \exp(i\beta_p L)
 \end{array} \right] .
\label{14}
\end{equation}
It also holds that $ \delta_p = 2\beta_p $, $ \Delta_p = \beta_p $
and $ D_p = \exp(i\beta_p L)/(2\beta_p) $ in this limit.

\section{Perturbation solution}

Perturbation solution of two equations (\ref{8}) can be found
after substituting the solution for the fundamental field
contained in Eq.~(\ref{12}). In the perturbation approach, the
inquired solution $ \hat{A}_{s_b}(z,\omega_s) $ for $ b=F,B $ is
expressed as $ \sum_{n=0}^{\infty} \hat{A}_{s_b}^{(n)}(z,\omega_s)
$ where an $ n $-th term $ \hat{A}_{s_b}^{(n)}(z,\omega_s) $ is
proportional to $ K_{nl,q}^n $.

\subsection{General solution up to the first order in nonlinearity}

The zeroth-order terms $ \hat{A}_{s_F}^{(0)}(z,\omega_s) $ and $
\hat{A}_{s_B}^{(0)}(z,\omega_s) $ are given as a solution to the
equations
\begin{eqnarray}    
 \frac{d\hat{A}_{s_F}^{(0)}(z,\omega_s)}{dz} &=&
  i\frac{\delta_{s}(\omega_s)}{2} \hat{A}_{s_F}^{(0)}(z,\omega_s) + iK_s(\omega_s)
  \hat{A}_{s_B}^{(0)}(z,\omega_s) , \nonumber \\
 \frac{d\hat{A}_{s_B}^{(0)}(z,\omega_s)}{dz} &=&
  - i\frac{\delta_{s}(\omega_s)}{2} \hat{A}_{s_B}^{(0)}(z,\omega_s)
  - iK_{s}^*(\omega_s) \hat{A}_{s_F}^{(0)}(z,\omega_s) . \nonumber\\
 & &
\label{15}
\end{eqnarray}
This solution can be written in the compact matrix form:
\begin{eqnarray} 
 \left[ \begin{array}{c} \hat{A}_{s_F}(z,\omega_s) \cr
  \hat{A}_{s_B}(z,\omega_s) \end{array} \right] &=&
  \sum_{\pm} \tilde{B}_s^{\pm}(\omega_s) \exp[\pm
  i\Delta_s(\omega_s)z]  \nonumber \\
 & & \mbox{} \times \left[ \begin{array}{c} \hat{A}_{s_F}(0,\omega_s) \cr
  \hat{A}_{s_B}(0,\omega_s) \end{array} \right];
\label{16}
\end{eqnarray}
$ \Delta_s(\omega_s) =
\sqrt{\delta_s^2(\omega_s)/4-|K_s(\omega_s)|^2} $. The matrices $
\tilde{B}_s^{\pm}(\omega_s) $ are expressed as
\begin{equation} 
 \tilde{B}_s^{\pm}(\omega_s) = \frac{1}{2\Delta_s}
  \left[ \begin{array}{cc} \pm\frac{\delta_s}{2}+\Delta_s
   & \pm K_s \cr
  \mp K_s^* & \mp \frac{\delta_s}{2}+\Delta_s
  \end{array} \right] .
\label{17}
\end{equation}

The equations for the first-order terms $
\hat{A}_{s_F}^{(1)}(z,\omega_s) $ and $
\hat{A}_{s_B}^{(1)}(z,\omega_s) $ have a more complex structure:
\begin{eqnarray}    
 \frac{d\hat{A}_{s_F}^{(1)}(z,\omega_s)}{dz} &=&
  i\frac{\delta_{s}(\omega_s)}{2} \hat{A}_{s_F}^{(1)}(z,\omega_s) + iK_s(\omega_s)
  \hat{A}_{s_B}^{(1)}(z,\omega_s) \nonumber \\
  & & \hspace{-15mm} \mbox{} + 4\int_{0}^{\infty}d\omega_s^{'} K_{nl,q}(\omega_s,\omega_s^{'})
  \exp[i\delta_{nl,q} z] \nonumber \\
  & & \hspace{-5mm} \mbox{} \times
  \hat{A}_{p_F}(z,\omega_s+\omega_s^{'}) \hat{A}^{(0)\dagger}_{s_F}(z,\omega_s^{'}) ,
  \nonumber\\
 \frac{d\hat{A}_{s_B}^{(1)}(z,\omega_s)}{dz} &=&
  - i\frac{\delta_{s}(\omega_s)}{2} \hat{A}_{s_B}^{(1)}(z,\omega_s)
  - iK_{s}^*(\omega_s) \hat{A}_{s_F}^{(1)}(z,\omega_s)
  \nonumber \\
  & & \hspace{-15mm} \mbox{} - 4\int_{0}^{\infty}d\omega_s^{'} K_{nl,q}(\omega_s,\omega_s^{'})
  \exp[-i\delta_{nl,q} z] \nonumber \\
  & & \hspace{-5mm} \mbox{} \times
  \hat{A}_{p_B}(z,\omega_s+\omega_s^{'})
  \hat{A}^{(0)\dagger}_{s_B}(z,\omega_s^{'}).
\label{18}
\end{eqnarray}

Equations (\ref{18}) for a fixed frequency $ \omega_s $ represent
a coupled set of two linear differential equations with nonzero
right-hand sides. They can be formally written as follows
\begin{equation}  
 \frac{dA(z)}{dz} = i K A(z) + F(z) ,
\label{19}
\end{equation}
where $ A $ and $ F $ are vectors with 2 elements and $ K $ is a $
2\times 2$ matrix. Direct inspection confirms a non-homogeneous
solution to Eq.~(\ref{19}) in the form
\begin{equation} 
 A(z) = \int_{0}^{z} dz' A^{\rm hom}(z-z')F(z')
\label{20}
\end{equation}
provided that $ dA^{\rm hom}(z)/(dz) = iK A^{\rm hom}(z) $.

The sum of the zeroth- and first-order solutions gives us the
solution to the equations (\ref{8}) valid up to $ K_{nl,q} $. It
can be again written in the matrix form using the matrices $
\tilde{B}_s^{\pm}(\omega_s) $ from Eq.~(\ref{17}):
\begin{eqnarray} 
 \left[ \begin{array}{c} \hat{A}_{s_F}(z,\omega_s) \cr
  \hat{A}_{s_B}(z,\omega_s) \end{array} \right] &=&
  \sum_{\pm} \tilde{B}_s^{\pm}(\omega_s) \exp[\pm i\Delta_a(\omega_a)z]
  \nonumber \\
 & & \hspace{-10mm} \mbox{} \times \left[ \begin{array}{c} \hat{A}_{s_F}(0,\omega_s) \cr
  \hat{A}_{s_B}(0,\omega_s) \end{array} \right] +
  \left[ \begin{array}{c} \hat{F}_{s_F}(z,\omega_s) \cr
  \hat{F}_{s_B}(z,\omega_s) \end{array} \right] .
\label{21}
\end{eqnarray}
Operators $ \hat{F}_{s_b}(z,\omega_s) $ for $ b=F,B $ describe the
solution proportional to the first power of nonlinear constants $
K_{nl,q} $. They are expressed for $ z=L $ below in
Eq.~(\ref{23}).

Now we write the solution in Eq.~(\ref{21}) for $ z=L $ and
partially invert the obtained linear relations in order to find
the input-output relations of the waveguide. The result can be
expressed as
\begin{eqnarray} 
 \left[ \begin{array}{c} \hat{A}_{s_F}(L,\omega_s) \cr
  \hat{A}_{s_B}(0,\omega_s) \end{array} \right] &=&
  \nonumber \\
 & & \hspace{-24mm}  D_s \left[ \begin{array}{cc} 2\Delta_s & 2iK_s\sin(\Delta_s L) \cr
   2iK_s^*\sin(\Delta_s L) & 2\Delta_s \end{array} \right]  \left[ \begin{array}{c}
  \hat{A}_{s_F}(0,\omega_s) \cr
  \hat{A}_{s_B}(L,\omega_s) \end{array} \right] \nonumber \\
 & & \hspace{-24mm} \mbox{}  +
  \left[ \begin{array}{c} \hat{F}_{s_F}(L,\omega_s) - 2iK_s D_s \sin(\Delta_s L)
  \hat{F}_{s_B}(L,\omega_s) \cr
  -2\Delta_s D_s \hat{F}_{s_B}(L,\omega_s) \end{array} \right].
\label{22}
\end{eqnarray}
In Eq.~(\ref{22}), the definition $ D_s(\omega_s) =
[2\Delta_s\cos(\Delta_s L) - i\delta_s \sin(\Delta_s L)]^{-1} $
has been used. The expressions on the right-hand side of
Eq.~(\ref{22}) can be substantially simplified if the signal field
fulfils the resonance condition $ \Delta_s \approx m \pi/L $, $
m=1,2,\ldots $, that gives the maximum enhancement of its
electric-field amplitudes in the waveguide. In this case, $
\sin(\Delta_s L) \approx 0 $.

The operators $ \hat{F}_{s_b}(L,\omega_s) $ at $ z=L $ are
obtained in the form:
\begin{eqnarray}   
 \hat{F}_{s_a}(L,\omega_s) &=& (\pm 1)_a 4 \int_{0}^{\infty}
  d\omega'_s K_{nl,q}(\omega_s,\omega'_s) \nonumber \\
 & & \hspace{-22mm} \mbox{} \times  \exp\left[ (\pm 1)_a
  \frac{i\delta_{nl,q}L}{2} \right] \sum_{\pm_s} \sum_{b=F,B}
  \tilde{B}_{s,ab}^{\pm_s} \exp\left[ \pm_s
  \frac{i\Delta_s(\omega_s)L}{2} \right] \nonumber \\
 & & \hspace{-22mm} \mbox{} \times
  \sum_{\pm_p} \sum_{c=F,B} B_{p,bc}^{\pm_p} \exp\left[ \pm_p
  \frac{i\Delta_p(\omega_s+\omega'_s)L}{2} \right]
  \hat{A}_{pc}(\omega_s+\omega'_s) \nonumber \\
 & & \hspace{-22mm} \mbox{}  \times
  \sum_{d=F,B} B_{s,bd}^{\pm_s*} \exp\left[ \mp_s
  \frac{i\Delta_s(\omega'_s)L}{2} \right]
  \hat{A}_{sd}^\dagger(\omega'_s) S_a^{\pm_p,\mp_s}(\omega_s,\omega'_s),
  \nonumber \\
 & & \hspace{2cm} a=F,B,
\label{23}
\end{eqnarray}
where
\begin{eqnarray}     
 S_a^{\pm_p,\pm_s}(\omega_s,\omega'_s) &=& 2L {\rm sinc} \bigl[
  [(\pm 1)_a\delta_{nl,q} \pm_p \Delta_p(\omega_s+\omega'_s) \nonumber \\
 & &  \hspace{-20mm} \mbox{}
  \pm_s \Delta_s(\omega_s) \pm_s \Delta_s(\omega'_s)]L/2 \bigr],
  \hspace{5mm} a=F,B,
\label{24}
\end{eqnarray}
$ {\rm sinc}(x) = \sin(x)/x $. The matrices $
B_{s}^{\pm}(\omega_s) $ introduced in Eq.~(\ref{23}) are defined
analogously to those given in Eq.~(\ref{13}) and characterizing
the fundamental field. We note that the expression in
Eq.~(\ref{23}) contains only one half of all possible terms. The
missing terms are far from the quasi-phase-matching conditions and
thus give negligible contributions.

\subsection{Gaussian spectral approximation in a resonant term}

The right-hand-side of Eq.~(\ref{23}) giving $
\hat{F}_{s_a}(L,\omega_s) $ is composed of four terms that differ
in the signs of eigenvalues $ \Delta_p $ and $ \Delta_s $
(resolved by the symbols $ \pm_p $, $ \pm_s $).
Quasi-phase-matching conditions in the waveguide are such that
they emphasize only one out of the four terms. The remaining terms
give small contributions. That is why we pay attention only to one
of them. We also consider only the incident forward-propagating
fundamental field in a multi-mode coherent state with a Gaussian
spectral shape and spectral phase variations such that the
fundamental field in the middle of the waveguide has the same
phase along the spectrum:
\begin{eqnarray}   
 A_{p_F}(0,\omega_p) &=& \xi_{p}
  \sqrt{ \frac{\tau_p}{\sqrt{2\pi}^3} } \exp\left[
  - \frac{\tau_p^2 (\omega_p-\omega_p^0)^2}{4} \right]
  \nonumber \\
 & & \mbox{} \times
   \exp\left[ \mp_p \frac{i\Delta_p(\omega_p)L}{2} \right],
  \nonumber \\
 A_{p_B}(L,\omega_p) &=& 0.
\label{25}
\end{eqnarray}
The fundamental pulse has amplitude $ \xi_p $, duration $ \tau_p $
and carrying frequency $ \omega^0_p $.

We further assume that both the fundamental and SSF spectra are
not too wide and so the propagation constants $ \beta_a(\omega_a)
$ can be approximated by their second-order Taylor expansions:
\begin{eqnarray}  
 \beta_a(\omega_a) &=& \beta_a^0 + \beta_{1a} (\omega_a-\omega_a^0)
  + \beta_{2a} (\omega_a-\omega_a^0)^2 , \nonumber \\
 \beta_a^0 &=& \beta_a(\omega_a^0) , \nonumber \\
  \beta_{ia} &=& \left. \frac{1}{i!} \frac{d^i \beta_a}{d\omega_a^i}
  \right|_{\omega_a=\omega_a^0} , \hspace{3mm} i=1,2; \hspace{3mm} a=p,s .
\label{26}
\end{eqnarray}
In this approximation, the eigenvalues $ \Delta_a(\omega_a) $ can
be expressed as
\begin{eqnarray}  
 \Delta_a(\omega_a) &=& \Delta_a^0 + \Delta_{1a} (\omega_a-\omega_a^0)
  + \Delta_{2a} (\omega_a-\omega_a^0)^2 , \nonumber \\
 \Delta_a^0 &=& \Delta_a(\omega_a^0) , \nonumber \\
 \Delta_{1a} &=& \frac{\beta_{1a}(\beta_a^0 -\pi/\Lambda_a)}{\Delta_a^0} ,
  \nonumber \\
 \Delta_{2a} &=& \frac{\beta_{2a}(\beta_a^0 -\pi/\Lambda_a)+\beta_{1a}^2}{2\Delta_a^0}
  - \frac{\beta_{1a}^2(\beta_a^0 -\pi/\Lambda_a)^2}{2(\Delta_a^0)^3}  ,
  \nonumber \\
 & &  \hspace{3mm} a=p,s .
\label{27}
\end{eqnarray}

Efficient nonlinear interaction occurs if quasi-phase-matching
conditions for the central frequencies $ \omega_p^0 $ and $
\omega_s^0 = \omega_p^0/2 $ are fulfilled, i.e.
\begin{equation}  
 \delta_{nl,q} \pm_p \Delta_p(\omega_p^0) \mp_s
 2\Delta_s(\omega_s^0) =0.
\label{28}
\end{equation}
The function $ S_a^{\pm_p,\pm_s} $ in Eq.~(\ref{24}) can then be
rewritten as
\begin{eqnarray}  
 S^{\pm_p,\pm_s}(\omega_s,\omega'_s) &=& 2L
  {\rm sinc} \Biggl( \Biggl[ (\pm_p \Delta_{1p} \pm_s \Delta_{1s}) \nonumber \\
 & & \hspace{-25mm} \mbox{} \times (\omega_s+\omega'_s-\omega_p^0)
  \pm_s \frac{\Delta_{2s}(\omega_s-\omega'_s)^2}{2} \Biggr]
  \frac{L}{2} \Biggr)
\label{29}
\end{eqnarray}
assuming $ \Delta_{1a} \gg \Delta_{2a}\Delta\omega $ for $ a=p,s
$; $ \Delta \omega $ characterizes the fields' spectral width.

The use of Gaussian approximation to the $ \rm sinc $ function in
Eq.~(\ref{29}) [$ {\rm sinc}(\alpha x + \beta y^2) \approx
\exp(-\alpha^2 x^2/5 - |\beta|y^2/3)] $ for constants $ \alpha $,
$ \beta $] allows to derive the useful relation:
\begin{eqnarray}  
 S^{\pm_p,\pm_s}(\omega_s,\omega'_s)A_{p_F}(0,\omega_s+\omega'_s) &=&
  2L \xi_{p} \sqrt{ \frac{\tau_p}{\sqrt{2\pi}^3} } \nonumber \\
 & & \mbox{} \hspace{-45mm}
  \times \exp\left[ \mp_p \frac{i\Delta_p(\omega_s+\omega'_s)L}{2} \right]
  \Phi(\omega_s,\omega_s^{'}) .
\label{30}
\end{eqnarray}
In Eq.~(\ref{30}), the function $ \Phi $,
\begin{eqnarray}  
 \Phi(\omega_s,\omega_s^{'}) &=&
  \exp\left[ - \frac{(\omega_s-\omega_s^{'})^2}{\Delta_{-}^2 }
   - \frac{(\omega_s+\omega_s^{'}-\omega_p^0)^2 }{\Delta_{+}^2 }
  \right] , \nonumber \\
 & &
\label{31} \\
 \frac{1}{\Delta_{-}^2} &=& \frac{\Delta_{2s}L}{12} , \nonumber \\
 \frac{1}{\Delta_{+}^2} &=& \frac{(\pm_p \Delta_{1p} \pm_s \Delta_{1s})^2
  L^2}{20} + \frac{\tau_p^2}{4} ,
 \label{32}
\end{eqnarray}
determines the structure of spectral modes.

These spectral modes can be revealed using the Schmidt
decomposition \cite{Luks2007x,Wasilewski2006} of function $ \Phi
$:
\begin{equation}   
 \Phi(\omega_s,\omega_s^{'}) = \sum_{n=0}^{\infty}
  \mu_n \phi_n(\omega_s) \phi_n(\omega_s^{'}) .
\label{33}
\end{equation}
Eigenvalues $ \mu_n $ introduced in Eq.~(\ref{33}) take the form:
\begin{eqnarray}   
 \mu_n &=& 2 \sqrt{ \frac{\pi}{\Delta_{+}\Delta_{-}}} \theta^{n/2} ,
\label{34} \\
 \theta &=& \left( \frac{\Delta_{+}-\Delta_{-}}{\Delta_{+}+\Delta_{-}} \right)^2 .
  \nonumber
\end{eqnarray}
Eigenmode spectral functions $ \phi_n $ can be expressed in terms
of the Hermite polynomials $ {\rm H_n} $:
\begin{eqnarray}    
 \phi_n(\omega_s) &=& \sqrt{ \frac{\tau_s}{2^n n!
  \sqrt{\pi} }} \exp \left[ - \tau_s^2(\omega_s-\omega_s^0)^2
  \right]  \nonumber \\
 & & \mbox{} \times
  {\rm H}_n(\tau_s[\omega_s-\omega_s^0]) ,
 \label{35} \\
 \tau_s &=& \sqrt{\frac{1-\theta^2}{\theta}} .
 \label{36}
\end{eqnarray}
For the considered waveguide, $ \Delta_{+} \ll \Delta_{-} $ and
thus a typical time constant $ \tau_s $ of the SSF field defined
in Eq.~(\ref{36}) can be approximated as:
\begin{equation}   
 \tau_s \approx 8 \frac{\Delta_{+}}{\Delta_{-}} .
\end{equation}

The Schmidt decomposition in Eq.~(\ref{33}) allows us to transform
Eqs.~(\ref{22}) written for the 'continuous index' $ \omega_s $ to
those related to spectral modes. The appropriate unitary
transformation of the input and output operators takes the form:
\begin{eqnarray}  
 \hat{a}_{s_F,n}^{\rm in} &=& \int_{0}^{\infty}
  d\omega_s \phi_n(\omega_s) \hat{a}_{s_F}(0,\omega_s)
  \exp\left[-\frac{i\pi L}{2\Lambda_s} \right] \nonumber \\
 & & \hspace{5mm} \mbox{} \times
  \exp\left[\mp_s \frac{i\Delta_s(\omega_s) L}{2} \right] ,
  \nonumber \\
 \hat{a}_{s_B,n}^{\rm in} &=& \int_{0}^{\infty}
  d\omega_s \phi_n(\omega_s) \hat{a}_{s_B}(L,\omega_s)
  \exp\left[\frac{i\pi L}{2\Lambda_s} \right] \nonumber \\
 & & \hspace{5mm} \mbox{} \times
  \exp\left[\pm_s \frac{i\Delta_s(\omega_s) L}{2} \right] ,
  \nonumber \\
 \hat{a}_{s_F,n}^{\rm out} &=& \int_{0}^{\infty}
  d\omega_s \phi_n(\omega_s) \hat{a}_{s_F}(L,\omega_s)
  \exp\left[-\frac{i\pi L}{2\Lambda_s} \right] \nonumber \\
 & & \hspace{5mm} \mbox{} \times
  \exp\left[\mp_s \frac{i\Delta_s(\omega_s) L}{2} \right] ,
  \nonumber \\
 \hat{a}_{s_B,n}^{\rm out} &=& \int_{0}^{\infty}
  d\omega_s \phi_n(\omega_s) \hat{a}_{s_B}(0,\omega_s)
  \exp\left[\frac{i\pi L}{2\Lambda_s} \right] \nonumber \\
 & & \hspace{5mm} \mbox{} \times
  \exp\left[\pm_s \frac{i\Delta_s(\omega_s) L}{2} \right] .
\end{eqnarray}

When weak spectral dependencies of multiplicative factors
occurring in Eqs.~(\ref{22}) and (\ref{23}) are neglected, the
transformed Eqs.~(\ref{22}) valid either close to the resonance
condition or for non-scattered SSF field are expressed as:
\begin{eqnarray} 
 \left[ \begin{array}{c} \hat{a}_{s_F,n}^{\rm out} \cr
  \hat{a}_{s_B,n}^{\rm out} \end{array} \right] &=&
   2\Delta_s^0 D_s^0 \exp\left[\frac{i\pi L}{\Lambda_s} \right]
   \left[ \begin{array}{cc} 1 & 0 \cr 0 & 1 \end{array} \right]
   \left[ \begin{array}{c}
  \hat{a}_{s_F,n}^{\rm in} \cr
  \hat{a}_{s_B,n}^{\rm in} \end{array} \right] \nonumber \\
 & & \mbox{}  +
  \left[ \begin{array}{c} \hat{F}_{s_F,n} \cr
  \hat{F}_{s_B,n} \end{array} \right] ;
\label{39}
\end{eqnarray}
$ D_s^0 = D_s(\omega_s^0) $.

Defining suitable nonlinear coupling constants $ \xi_n $ for
individual spectral modes,
\begin{equation}   
 \xi_n = 8LK_{nl,q}(\omega_s^0,\omega_s^0) \sqrt{\frac{\tau_p}{\sqrt{2\pi}^3}}
  \xi_{p} \mu_n,
\label{40}
\end{equation}
the operator coefficients $ \hat{F}_{s_F,n} $ and $
\hat{F}_{s_B,n} $ introduced in Eq.~(\ref{39}) take the form:
\begin{eqnarray}  
 \hat{F}_{s_F,n} &=& \xi_n
  \exp\left[\frac{i\delta_{nl,q}L}{2}\right]
  \frac{D_p^0 D_s^{0*}}{2\Delta_s^0} \Biggl\{ \left[ \left( \pm_s\frac{\delta_s^0}{2}+
  \Delta_s^0\right)^2 \right. \nonumber \\
 & & \hspace{-15mm} \left. \mbox{} \times
  \left( \pm_p\frac{\delta_p^0}{2}+\Delta_p^0\right) \pm_p
  K_p^{0*} (K_s^0)^2 \right]  \hat{a}_{s_F,n}^{{\rm in}\dagger}
  + \left[ \left(
   \frac{\delta_s^0}{2}\pm_s\Delta_s^0\right) \right.    \nonumber \\
 & & \hspace{-15mm} \left. \mbox{} \times
  K_s^{0*} \left( \pm_p\frac{\delta_p^0}{2}+\Delta_p^0\right) \pm_p
  K_s^0 K_p^{0*} \left( \frac{\delta_s^0}{2} \mp_s \Delta_s^0\right)
  \right]  \hat{a}_{s_B,n}^{{\rm in}\dagger} \Biggr\} ,
  \nonumber \\
 \hat{F}_{s_B,n} &=& - \xi_n
  \exp\left[-\frac{i\delta_{nl,q}L}{2}\right]
  \exp\left(\pm_s i\Delta_s^0 L \right)
   D_p^0 |D_s^{0}|^2  \nonumber \\
 & & \hspace{-10mm} \mbox{} \times \Biggl\{ \left[ K_s^{0*} \left( \frac{\delta_s^0}{2}
  \pm_s \Delta_s^0\right) \left( \mp_p\frac{\delta_p^0}{2}-\Delta_p^0\right) \mp_p
  K_p^{0*} K_s^{0*} \right. \nonumber \\
 & &  \hspace{-10mm}  \left. \mbox{} \times \left( \frac{\delta_s^0}{2}
  \mp_s \Delta_s^0\right) \right]  \hat{a}_{s_F,n}^{{\rm in}\dagger}
  + \left[ \left( \mp_p \frac{\delta_p^0}{2}-\Delta_p^0\right) (K_s^{0*})^2 \right.
   \nonumber \\
 & & \hspace{-10mm} \left. \mbox{} \mp_p K_p^{0*} \left( \frac{\delta_s^0}{2}
  \mp_s \Delta_s^0\right)^2 \right]  \hat{a}_{s_B,n}^{{\rm in}\dagger} \Biggr\};
\label{41}
\end{eqnarray}
$ D_p^0 = D_p(\omega_p^0) $.

If the corrugation is only in the fundamental field, the
expressions for operator coefficients $ \hat{F}_{s_F,n} $ and $
\hat{F}_{s_B,n} $ in Eq.~(\ref{41}) simplify:
\begin{eqnarray}  
 \hat{F}_{s_F,n} &=& \xi_n
 \exp\left[\frac{i\delta_{nl,q}L}{2}\right]
   D_p^0 \exp(-i\beta_s^0 L) \nonumber \\
 & & \mbox{} \times  \left( \pm_p\frac{\delta_p^0}{2}+\Delta_p^0\right)
   \hat{a}_{s_F,n}^{{\rm in}\dagger} , \nonumber \\
 \hat{F}_{s_B,n} &=& - \xi_n
 \exp\left[\frac{-i\delta_{nl,q}L}{2}\right] D_p^0 \exp(-i\beta_s^0 L)
   \nonumber \\
 & &  \mbox{} \times  (\mp_p K_p^{0*})\hat{a}_{s_B,n}^{{\rm in}\dagger} .
\label{42}
\end{eqnarray}

On the other hand, the corrugation resonating with the SSF field
leads to the following expressions
\begin{eqnarray}  
 \hat{F}_{s_F,n} &=& \xi_n
 \exp\left[\frac{i\delta_{nl,q}L}{2}\right]
  \exp(i\beta_p^0 L) \frac{D_s^{0*}}{2\Delta_s^0}
   \left( \pm_s\frac{\delta_s^0}{2}+\Delta_s^0\right) \nonumber \\
 & & \mbox{} \times \left\{ \left( \pm_s\frac{\delta_s^0}{2}+\Delta_s^0\right)
   \hat{a}_{s_F,n}^{{\rm in}\dagger}  \pm K_s^{0*} \hat{a}_{s_B,n}^{{\rm in}\dagger}
   \right\} , \nonumber \\
 \hat{F}_{s_B,n} &=& -\xi_n
  \exp\left[-\frac{i\delta_{nl,q}L}{2}\right]
  \exp(i\beta_p^0 L) \exp(\pm_s i\Delta_s^0 L)
   \nonumber \\
 & & \hspace{-13mm} \mbox{} \times |D_s^{0}|^2 K_s^{0*} \left\{ \left( -\frac{\delta_s^0}{2} \mp_s \Delta_s^0\right)
   \hat{a}_{s_F,n}^{{\rm in}\dagger}  - K_s^{0*} \hat{a}_{s_B,n}^{{\rm in}\dagger}
   \right\} .
\label{43}
\end{eqnarray}

Multiplicative factors occurring in the expressions in
Eqs.~(\ref{41}---\ref{43}) describe the enhancement of nonlinear
interaction due to scattering of the fundamental and SSF fields.
This enhancement can be quantified by the expression $
(\pm_a\delta_a^0/2 + \Delta_a^0) / (2\Delta_a^0) $ for $ a=p,s $.
Let us consider first the corrugation present in the fundamental
field. In this case, the quasi-phase-matching condition written in
Eq.~(\ref{28}) takes the form
\begin{equation}  
 \delta_{nl,q}^{{\rm nat},0} - \frac{\delta_p^0}{2} \pm_p \Delta_p^0 =0 ,
\label{44}
\end{equation}
where $ \delta_{nl,q}^{\rm nat,0} = \beta_p^0 - 2\beta_s^0 + 2\pi
q / \Lambda_{nl} $ describes the natural quasi-phase mismatch.
Efficient quasi-phase-matching can be reached only if $
\delta_{nl,q}^{{\rm nat},0} $ and $ \delta_p^0 $ have the same
sign \cite{PerinaJr2007a}. Considering a positive value of phase
mismatch $ \delta_{nl,q}^{\rm nat,0} $, the sign + [-] in
Eq.~(\ref{44}) is suitable for $ \delta_{nl,q}^{\rm nat,0} <
|K_p^0| $ [$ \delta_{nl,q}^{\rm nat,0} > |K_p^0| $]. The
enhancement of fundamental-field amplitudes is then described by
the expression
\begin{equation}  
 \frac{\delta_p^0/2 + \Delta_p^0}{2\Delta_p^0} = \frac{1\pm_p 1}{2} +
 \frac{\delta_{nl,q}^{\rm nat,0}}{2\Delta_p^0} > 1.
\label{45}
\end{equation}
On the other hand, a negative value of phase mismatch $
\delta_{nl,q}^{\rm nat,0} $ requires the opposite choice of signs
in Eq.~(\ref{44}) and the enhancement of fundamental-field
amplitudes can be quantified by the expression
\begin{equation}  
 \frac{-\delta_p^0/2 + \Delta_p^0}{2\Delta_p^0} = \frac{1\mp_p 1}{2}
 - \frac{\delta_{nl,q}^{\rm nat,0}}{2\Delta_p^0} > 1.
\label{46}
\end{equation}
The presence of corrugation in the SSF field needs the following
quasi-phase-matching conditions:
\begin{equation}  
 \delta_{nl,q}^{\rm nat,0} + \delta_s^0 \mp_s 2\Delta_s^0 =0.
\label{47}
\end{equation}
In this case, the quantities $ \delta_{nl,q}^{\rm nat,0} $ and $
\delta_s^0 $ have to differ in their signs. The enhancement
factors of the SSF-field amplitudes can be analyzed similarly as
for the fundamental field.

\subsection{Principal squeeze variance}

The enhancement of electric-field amplitudes due to the presence
of the corrugation results in larger squeezing of fluctuations of
these amplitudes. The suppression of amplitude fluctuations can be
quantified by a principal squeeze variance $ \lambda $
\cite{Luks1988,Perina1991,PerinaJr2000} that can be determined
along the relations
\begin{eqnarray}   
 \lambda_{s_b} &=& 1+ 2\left( B_{s_b}-|C_{s_b}|\right),
\label{48} \\
 B_{s_b} &=& \langle \Delta\hat{a}^\dagger_{s_b}
  \Delta\hat{a}_{s_b} \rangle ,  \nonumber  \\
 C_{s_b} &=& \langle (\Delta\hat{a}_{s_b})^2 \rangle ,
  \hspace{10mm} b=F,B.
\label{49}
\end{eqnarray}
In Eqs.~(\ref{49}), $ \Delta\hat{a} = \hat{a} - \langle \hat{a}
\rangle $ and symbol $ \langle \;\; \rangle $ means the quantum
mechanical mean value.

Substitution of the expressions in Eq.~(\ref{41}) into
Eqs.~(\ref{48}) and (\ref{49}) provides the formulas for principal
squeeze variances $ \lambda_{s_F,n} $ and $ \lambda_{s_B,n} $ of
individual spectral modes,
\begin{eqnarray}   
 \lambda_{s_F,n} &=& 1 - \frac{|\xi_n|}{4\Delta_p^0 (\Delta_s^0)^2}
  \left| \left( \pm_s \frac{\delta_s^0}{2} + \Delta_s^0 \right)
  \left( \pm_p \frac{\delta_p^0}{2} + \Delta_p^0 \right) \right.
   \nonumber \\
 & & \hspace{-17mm}  \left. \mbox{} \times
  \left( \pm_s \frac{\delta_s^0}{2} + \Delta_s^0 \pm K_s^{0*} \right)
  \pm_p K_p^{0*} K_s^0
  \left( \frac{\delta_s^0}{2} \mp_s \Delta_s^0 + K_s^{0}
  \right) \right| , \nonumber \\
 \lambda_{s_B,n} &=& 1 - \frac{|\xi_n|}{4\Delta_p^0 (\Delta_s^0)^2}
  \left| K_s^{0*}
  \left( \pm_p \frac{\delta_p^0}{2} + \Delta_p^0 \right) \right.
   \nonumber \\
 & & \hspace{-7mm} \mbox{} \times
  \left( -\frac{\delta_s^0}{2} \mp_s \Delta_s^0 - K_s^{0*} \right)
  \mp_p K_p^{0*} \left( \mp_s\frac{\delta_s^0}{2} + \Delta_s^0 \right)
  \nonumber \\
 & & \hspace{-7mm}  \left. \mbox{} \times
  \left( \mp_s\frac{\delta_s^0}{2} + \Delta_s^0 \mp_s K_s^{0}
  \right) \right| .
\label{50}
\end{eqnarray}
The enhancement factors discussed for the fundamental field in
Eqs.~(\ref{45}) and (\ref{46}) are clearly recognized in the
expressions~(\ref{50}). Also the factors $ \delta_s^0/2 \pm_s
\Delta_s^0 + K_s^{0} $ and $ \delta_s^0/2 \pm_s \Delta_s^0 +
K_s^{0*} $ found in Eqs.~(\ref{50}) significantly contribute to
the enhancement of squeezing of SSF electric-field amplitudes due
to the same signs in front of $ \delta_s^0 $ and $ K_s^0 $.

The formulas in Eq.~(\ref{50}) considerably simplify for the
corrugation in the fundamental field only:
\begin{eqnarray}   
 \lambda_{s_F,n} &=& 1 - \frac{|\xi_n|}{\Delta_p^0 } \left|
  \left( \pm_p \frac{\delta_p^0}{2} + \Delta_p^0 \right) \right|,
   \nonumber \\
 \lambda_{s_B,n} &=& 1 - \frac{|\xi_n K_p^0|}{\Delta_p^0 } .
\label{51}
\end{eqnarray}

On the other hand, the presence of corrugation only in the SSF
field leaves us with the expressions
\begin{eqnarray}   
 \lambda_{s_F,n} &=& 1 - \frac{|\xi_n|}{2(\Delta_s^0)^2}
  \left| \left( \pm_s \frac{\delta_s^0}{2} + \Delta_s^0 \right)
  \right.  \nonumber \\
 & & \left. \mbox{} \times
  \left( \pm_s \frac{\delta_s^0}{2} + \Delta_s^0 \pm K_s^{0*} \right)
   \right| , \nonumber \\
 \lambda_{s_B,n} &=& 1 - \frac{|\xi_nK_s^0|}{2(\Delta_s^0)^2}
  \left| \left( \pm_s\frac{\delta_s^0}{2} + \Delta_s^0 \pm_s K_s^{0*}
  \right) \right| .
\label{52}
\end{eqnarray}

\section{Non-perturbation solution for non-dispersion propagation}

\begin{figure}    
 {\raisebox{3.3 cm}{a)} \hspace{5mm}
 \resizebox{0.7\hsize}{!}{\includegraphics{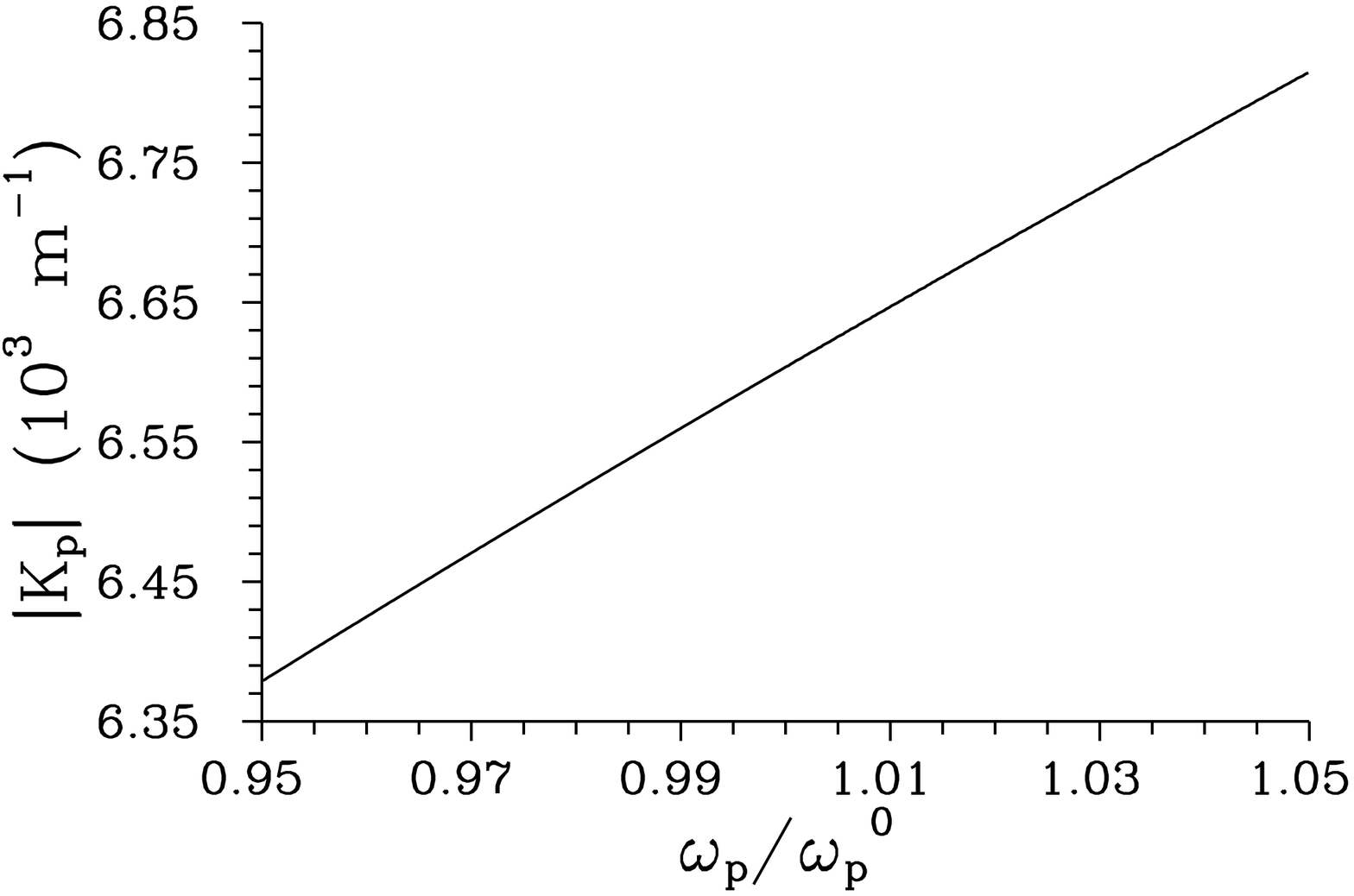}}

 \vspace{3mm}
 \raisebox{3.3 cm}{b)} \hspace{5mm}
 \resizebox{0.7\hsize}{!}{\includegraphics{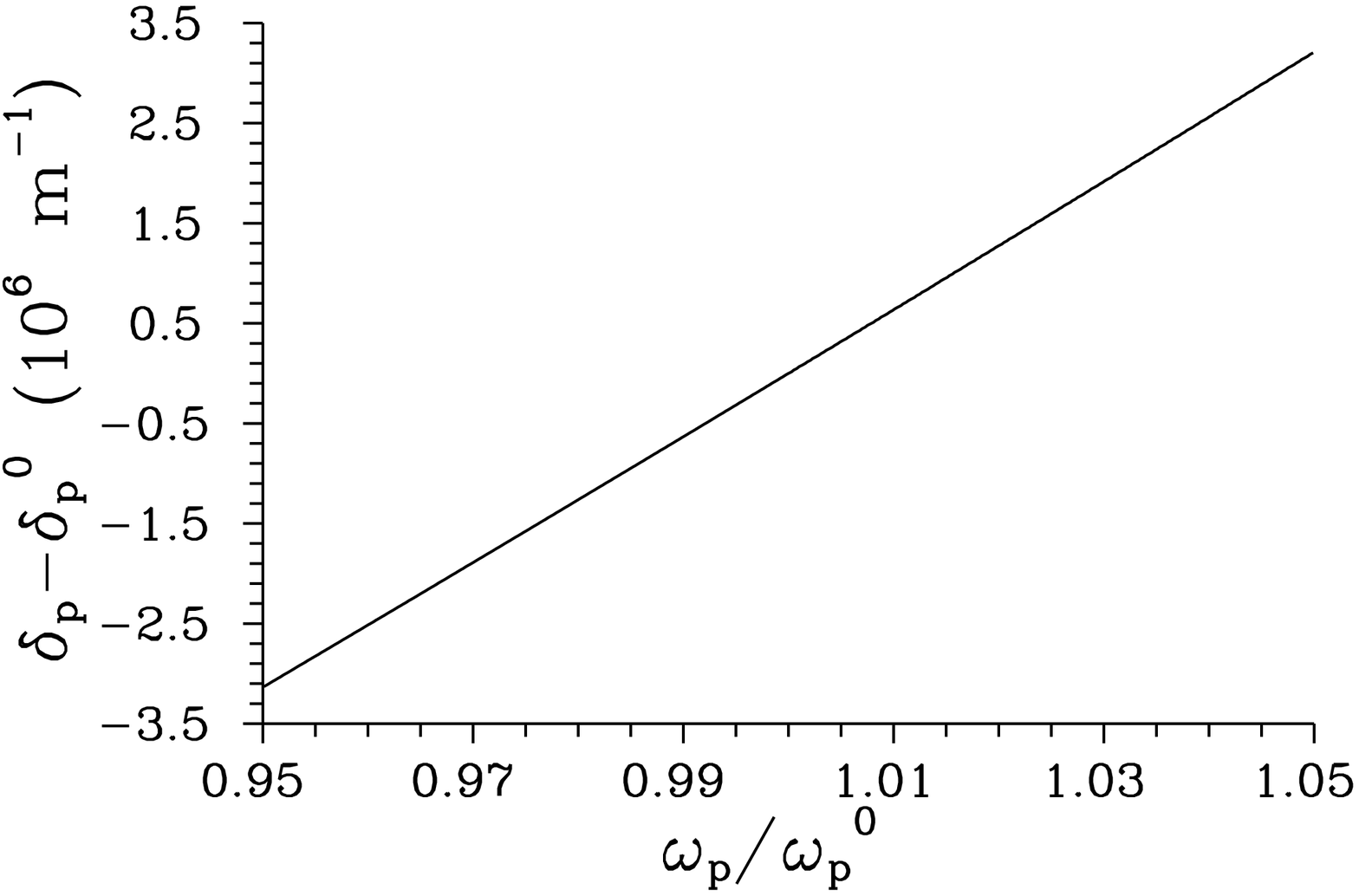}}}
 \vspace{0mm}

 \caption{(a) Absolute value of linear coupling
 constant $ K_p  $ and (b) linear phase mismatch $
 \delta_p - \delta_p^{0} $
 for the fundamental field as they depend on relative
 frequency $ \omega_p/\omega_p^0 $;
 $ t=5\times 10^{-7} $~m, $ t_l = 5 \times 10^{-8} $~m.}
\label{fig2}
\end{figure}

\begin{figure}    
 {\raisebox{3.3 cm}{a)} \hspace{5mm}
 \resizebox{0.7\hsize}{!}{\includegraphics{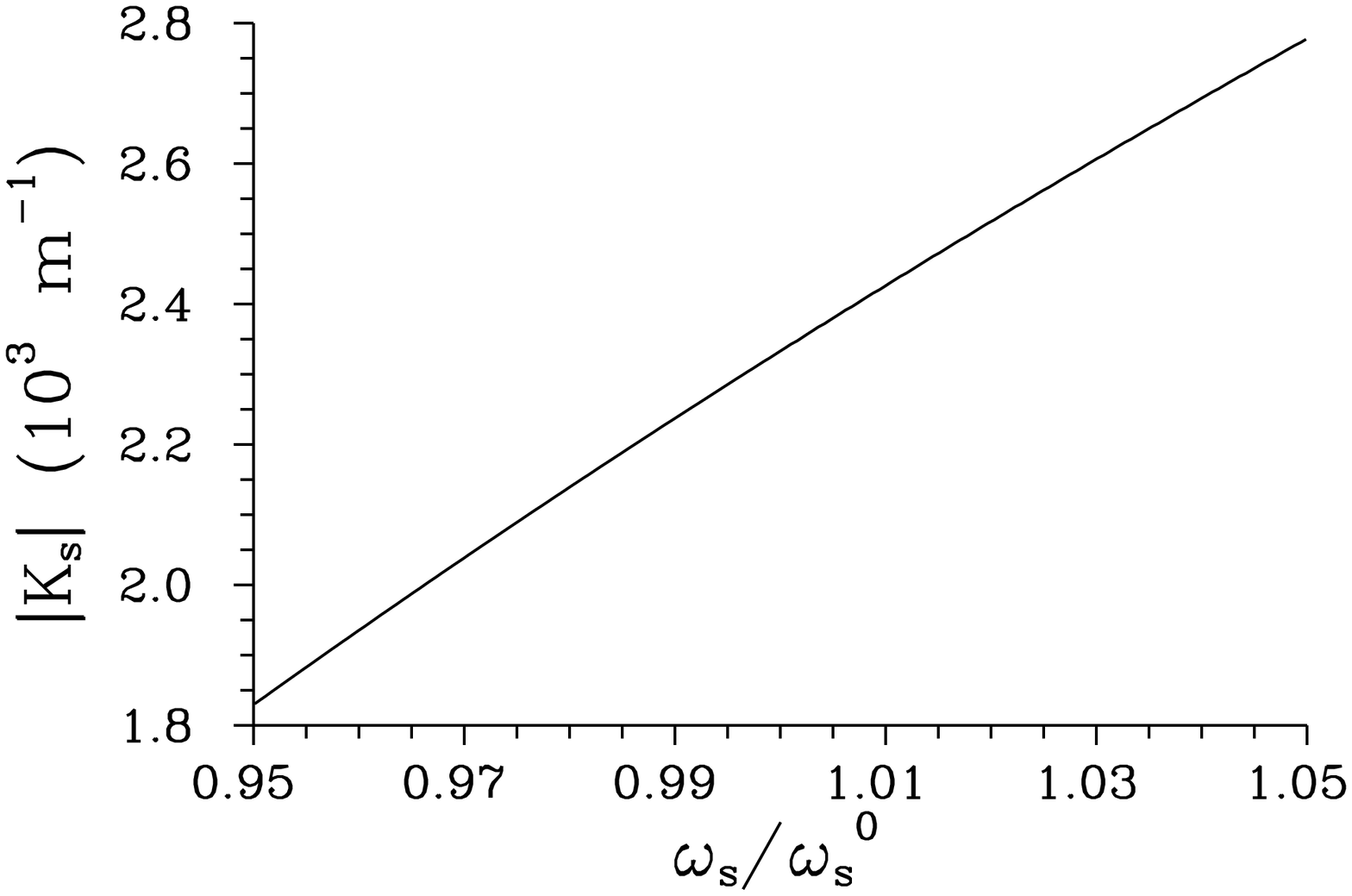}}

 \vspace{2mm}
 \raisebox{3.3 cm}{b)} \hspace{5mm}
 \resizebox{0.7\hsize}{!}{\includegraphics{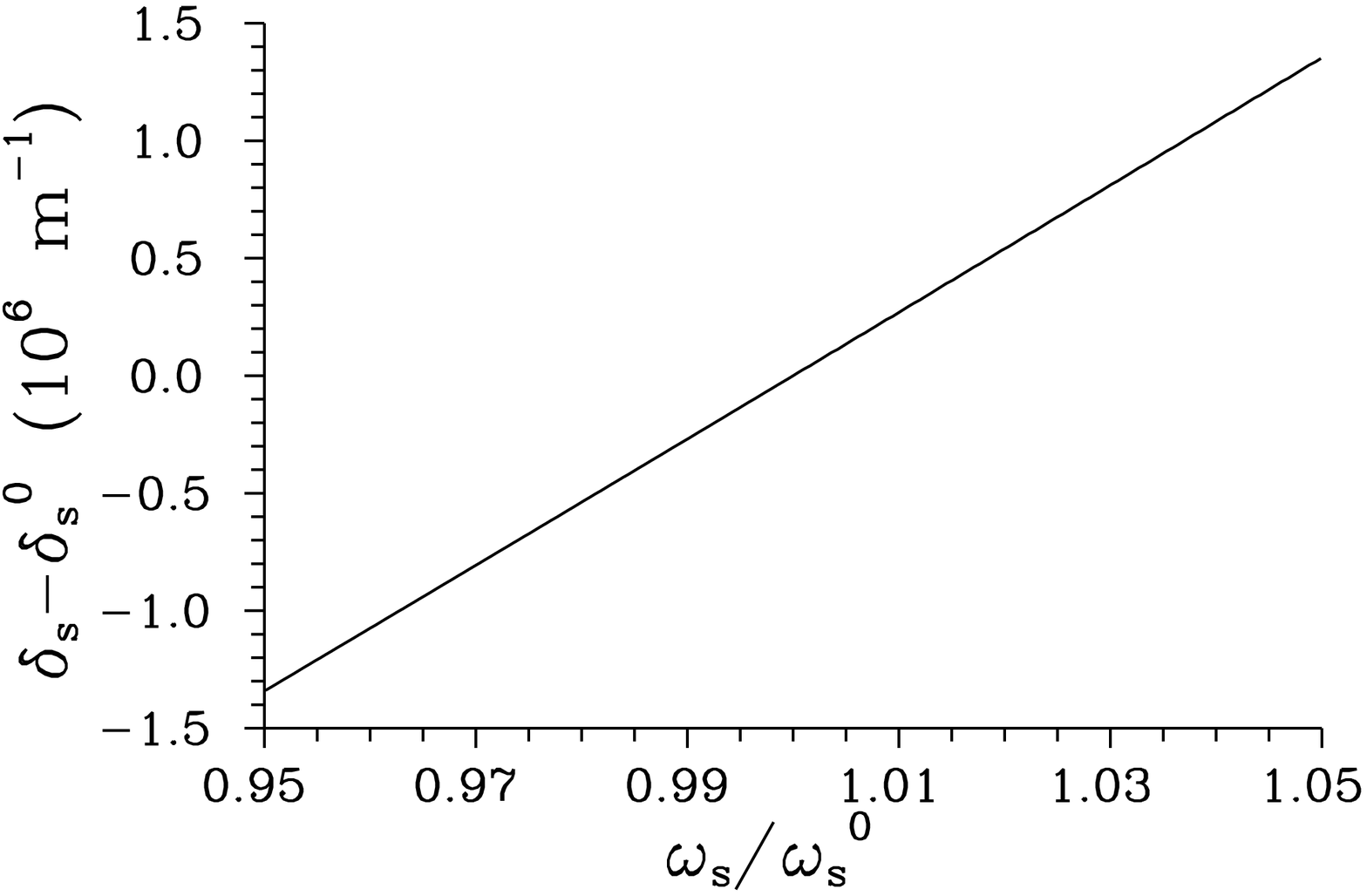}}}
 \vspace{0mm}

 \caption{(a) Absolute value of linear coupling
 constant $ K_s $ and (b) linear phase mismatch $
 \delta_s - \delta_s^{0} $ for the
 SSF field as functions of relative
 frequency $ \omega_s/\omega_p^0 $;
 values of parameters are the same as in Fig.~\ref{fig2}.}
\label{fig3}
\end{figure}

\begin{figure}    
 {\raisebox{4.3 cm}{a)} \hspace{4mm}
 \resizebox{.85\hsize}{!}{\includegraphics{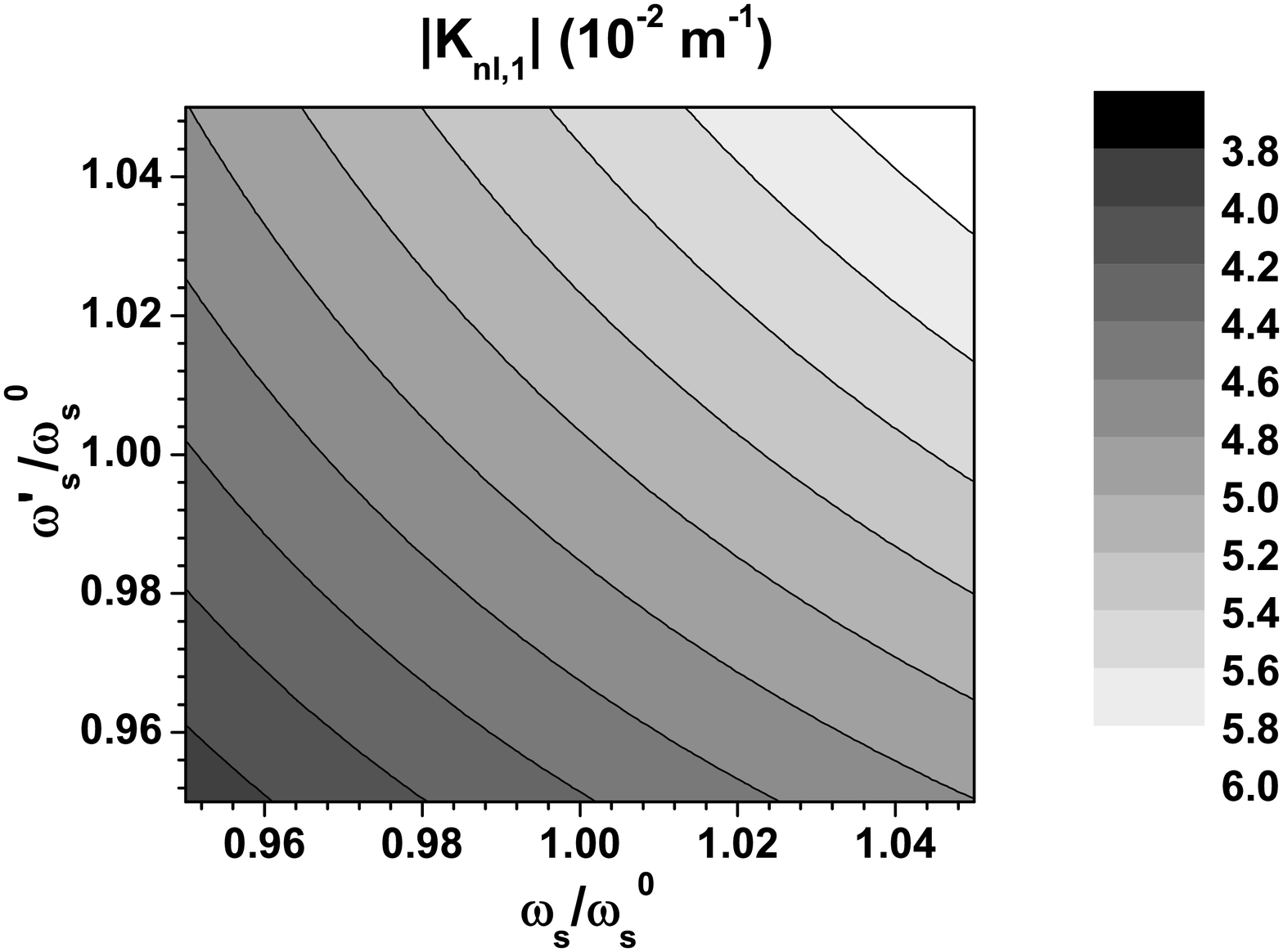}}

 \vspace{3mm}
 \raisebox{4.3 cm}{b)} \hspace{4mm}
 \resizebox{.85\hsize}{!}{\includegraphics{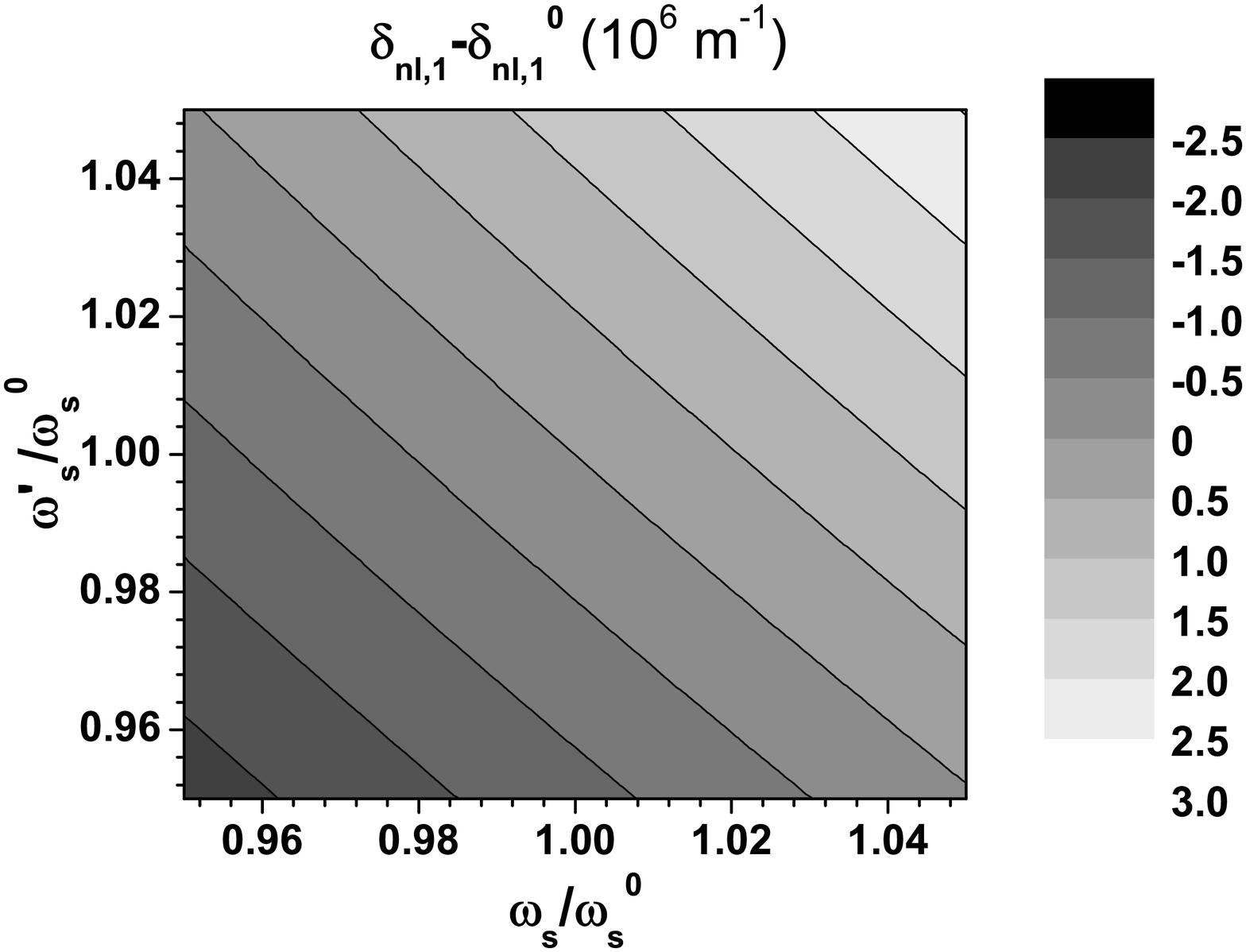}}}
 \vspace{0mm}

 \caption{Contour plots of (a) absolute value $ |K_{nl,1}| $ of nonlinear coupling
 constant and (b) nonlinear phase mismatch $
 \delta_{nl,1} - \delta_{nl,1}^{0} $ as they depend on relative
 frequencies $ \omega_s/\omega_p^0 $ and $ \omega_s^{'}/\omega_p^0
 $. Values of parameters are the same as in Fig.~\ref{fig2}.}
\label{fig4}
\end{figure}

The linear coupling constants $ K_p $, $ K_s $ and nonlinear
coupling constants $ K_{nl,q} $ depend usually only weakly on
frequencies in a relatively wide interval. Weak frequency
dependence of the coupling constants $ K_p $, $ K_s $ and $
K_{nl,1} $ in the considered range [approx. 80~nm (40~nm) for the
fundamental (SSF) field] is shown in Figs.~\ref{fig2}a,
\ref{fig3}a, and \ref{fig4}a, respectively, for the analyzed
waveguide. Also inter-mode dispersion both in the fundamental and
SSF fields can be in the first approximation neglected (see
Figs.~\ref{fig2}b, \ref{fig3}b, and \ref{fig4}b for frequency
dependence of the linear phase mismatches $ \delta_p $ and $
\delta_s $ and nonlinear phase mismatch $ \delta_{nl,1} $). We
note that the needed spectral range of the SSF field roughly
depends inversely proportionally on the waveguide length $ L $.
The solution of operator equations considerably simplifies when
the frequency dependencies are omitted and even certain analytical
results can be obtained.

In this case, Eqs.~(\ref{8}) and (\ref{9}) for the operator
amplitudes $ \hat{A}_{a_b}(z,\omega) $, $ a=p,s $, $ b=F,B $ can
be decoupled in their frequency 'index' using the Fourier
transform:
\begin{eqnarray}    
 \hat{A}_{a_b}(z,\tau_a) &=& \int_{-\infty}^{\infty}
  d\omega_a \hat{A}_{a_b}(z,\omega_a) \exp(-i\omega_a\tau_a) ,
\label{53} \\
 \hat{A}_{a_b}(z,\omega_a) &=& \frac{1}{2\pi} \int_{-\infty}^{\infty}
  d\tau_a \hat{A}_{a_b}(z,\tau_a) \exp(i\omega_a\tau_a) .
\label{54}
\end{eqnarray}
After the transformation, Eqs.~(\ref{8}) attain the form:
\begin{eqnarray}    
 \frac{d\hat{A}_{s_F}(z,\tau)}{dz} &=&
  i\frac{\delta_{s}^0}{2} \hat{A}_{s_F}(z,\tau) + iK_s^0
  \hat{A}_{s_B}(z,\tau) \nonumber \\
  & & \hspace{-18mm} \mbox{} + 4K_{nl,q}^0
  \exp[i\delta_{nl,q} z]
  \hat{A}_{p_F}(z,\tau) \hat{A}^\dagger_{s_F}(z,\tau) ,
  \nonumber\\
 \frac{d\hat{A}_{s_B}(z,\tau)}{dz} &=&
  - i\frac{\delta_{s}^0}{2} \hat{A}_{s_B}(z,\tau)
  - iK_{s}^{0*} \hat{A}_{s_F}(z,\tau)
  \nonumber \\
  & & \hspace{-18mm} \mbox{} - 4 K_{nl,q}^0
  \exp[-i\delta_{nl,q} z]
  \hat{A}_{p_B}(z,\tau) \hat{A}^\dagger_{s_B}(z,\tau) ;
\label{55}
\end{eqnarray}
$ K_{nl,q}^0 = K_{nl,q}(\omega_s^0,\omega_s^0) $. The solution of
Eqs.~(\ref{9}) written in Eq.~(\ref{12}) is transformed in the
considered approximation of strong fundamental field as follows:
\begin{eqnarray} 
 \left[ \begin{array}{c} \hat{A}_{p_F}(z,\tau) \cr
  \hat{A}_{p_B}(z,\tau) \end{array} \right] =
  \sum_{\pm} B_p^{\pm 0} \exp[\pm i\Delta_p^0 z]
  \left[ \begin{array}{c} \hat{A}_{p_F}(0,\tau) \cr
  \hat{A}_{p_B}(L,\tau) \end{array} \right]
\label{56}
\end{eqnarray}
and $ B_p^{\pm 0} = B_p^{\pm}(\omega_p^0) $ defined in
Eq.~(\ref{13}). The fundamental-field operator amplitudes $
\hat{A}_{p_F}(0,\tau) $ and $ \hat{A}_{p_B}(L,\tau) $ describe the
incident pulses.

We further discuss the solution to Eqs.~(\ref{55}) for the
corrugation present either in the fundamental or SSF field.

\subsection{Corrugation in the pump field only}

In this case, two operator equations (\ref{55}) are independent.
Moreover, we are interested in a solution close to the resonance
where the nonlinear terms in Eqs.~(\ref{55}) give considerable
contribution. We further pay attention to the SSF
forward-propagating field and assume only the incident fundamental
field at $ z=0 $. Using Eq.~(\ref{56}) the first equation in
(\ref{55}) can be rewritten for two different resonant conditions
indicated by upper indices $ \pm $,
\begin{eqnarray}   
 \frac{d\hat{A}_{s_F}(z,\tau)}{dz} &=& i\frac{\delta_{s}^0}{2} \hat{A}_{s_F}(z,\tau)
  + 4 K^{\pm}(\tau)
  \nonumber \\
 & & \hspace{-20mm} \mbox{} \times
  \exp[i(\delta_{nl,q}^0 \pm \Delta_p^0) z]
  \hat{A}^\dagger_{s_F}(z,\tau).
\label{57}
\end{eqnarray}
The effective nonlinear coupling constant $ K^{\pm}(\tau) \equiv
K_{nl,q}^0 B_{p,FF}^{\pm 0} A_{p_F}(0,\tau) $ incorporates the
enhancement of nonlinear interaction due to the fundamental-field
scattering.

The substitution $ \hat{A}_{s_F}(z,\tau) = \hat{\cal
A}_{s_F}(z,\tau) \exp[ i(\delta_{nl,q}^0 \pm \Delta_p^0) z/2] $, $
\hat{A}_{s_F}^\dagger(z,\tau) = \hat{\cal A}^\dagger_{s_F}(z,\tau)
\exp[ -i(\delta_{nl,q}^0 \pm \Delta_p^0) z/2] $ in Eq.~(\ref{57})
leads to differential equations with constant coefficients obeyed
by the operator amplitudes $ \hat{\cal A}_{s_F} $ and $ \hat{\cal
A}^\dagger_{s_F} $. Their solution transformed to the original
operators can be written as:
\begin{eqnarray}   
 \hat{A}_{s_F}(L,\tau) &=& U_{FF}(\tau) \hat{A}_{s_F}(0,\tau)
  + V_{FF}(\tau) \hat{A}_{s_F}^\dagger(0,\tau) , \nonumber\\
 & & 
 \label{58} \\
 U_{FF}(\tau) &=& \frac{1}{2} \exp(i\beta_s^0 L) \exp\left(i\frac{\Omega^{\pm} L}{2} \right)
  \nonumber \\
 & & \hspace{-5mm} \mbox{} \times
  \left[ \left( 1-\frac{i\Omega^{\pm}}{2\lambda^{\pm}(\tau)} \right)
  \exp[\lambda^{\pm}(\tau)L] \right. \nonumber \\
 & & \hspace{-5mm} \mbox{} \left. +
  \left( 1+\frac{i\Omega^{\pm}}{2\lambda^{\pm}(\tau)} \right)
  \exp[-\lambda^{\pm}(\tau)L] \right] ,
  \nonumber \\
 V_{FF}(\tau) &=& \frac{2K^{\pm}(\tau)}{\lambda^{\pm}(\tau)}
   \exp(i\beta_s^0 L)
  \exp\left(i\frac{\Omega^{\pm} L}{2} \right) \nonumber \\
 & & \hspace{-5mm} \mbox{} \times
  \left[ \exp[\lambda^{\pm}(\tau) L] - \exp[-\lambda^{\pm}(\tau) L]
  \right] .
 \label{59}
\end{eqnarray}
The phase mismatches $ \Omega^{\pm} $ and eigenvalues $
\lambda^{\pm} $ are given along the expressions:
\begin{eqnarray}    
 \Omega^{\pm} &=& \delta_{nl,q} - \delta_s^0 \pm \Delta_p^0 ,
  \nonumber \\
 \lambda^{\pm}(\tau) &=& \sqrt{ 16|K^{\pm}(\tau)|^2 -
  \Omega^{\pm 2}/4 } .
\label{60}
\end{eqnarray}
Ideal phase matching leads to $ \Omega^{\pm} = 0 $.

The inverse Fourier transform (\ref{54}) of the expression for
operator amplitude $ \hat{A}_{s_F}(L,\tau) $ in Eq.~(\ref{58})
provides the spectral operator amplitude $
\hat{A}_{s_F}(L,\omega_s) $ that equals the operator amplitude $
\hat{a}_{s_F}^{\rm out}(\omega_s) $ in the limit $ \Lambda_s
\longrightarrow \infty $,
\begin{eqnarray}    
 \hat{a}_{s_F}^{\rm out}(\omega_s) &=&
  \left[ \int_{0}^{\infty} d\omega_s^{'}
  U_{FF}(\omega_s-\omega_s^{'}) \hat{a}_{s_F}^{\rm in}(\omega_s^{'})
  \right. \nonumber \\
 & & \hspace{-5mm} \mbox{} \left. + \int_{0}^{\infty} d\omega_s^{'}
  V_{FF}(\omega_s+\omega_s^{'}) \hat{a}_{s_F}^{{\rm in}\dagger}(\omega_s^{'})
  \right] .
\label{61}
\end{eqnarray}
The functions $ U_{FF}(\omega) $ and $ V_{FF}(\omega) $ are given
by the inverse Fourier transform of the expressions in
Eqs.~(\ref{59}) and can be found numerically.

\subsection{Corrugation in the second-subharmonic field only}

We consider only the forward-propagating fundamental field and
quasi-phase-matching of the nonlinear interaction given by the
condition $ \delta_{nl,q} +\delta_p^0/2 = 0 $. The equations
(\ref{55}) can then be written in the following matrix form
\begin{eqnarray} 
 \frac{d}{dz} \left[ \begin{array}{c} \hat{A}_{s_F}(z,\tau) \cr
  \hat{A}_{s_B}(z,\tau) \cr \hat{A}_{s_F}^\dagger(z,\tau) \cr
  \hat{A}_{s_B}^\dagger(z,\tau)  \end{array} \right] &=&
  i M \left[ \begin{array}{c} \hat{A}_{s_F}(z,\tau) \cr
  \hat{A}_{s_B}(z,\tau) \cr \hat{A}_{s_F}^\dagger(z,\tau) \cr
  \hat{A}_{s_B}^\dagger(z,\tau)  \end{array} \right] , \nonumber \\
 & & \hspace{-30mm} M(\tau) = \left[ \begin{array}{cccc}
  \delta_s^0/2 & K_s^0 & -4iK_F & 0 \cr
  -K_s^{0*} & - \delta_s^0/2 & 0 & 0 \cr
  -4iK_F^* & 0 & -\delta_s^0/2 & -K_s^{0*} \cr
  0 & 0 & K_s^0 & \delta_s^0/2 \end{array} \right]
\label{62}
\end{eqnarray}
introducing the nonlinear coupling constant $ K_F(\tau) =
K_{nl,q}^0 A_{p_F}(0,\tau) $.

Eigenvalues $ \lambda_j $ of matrix $ M $ in Eq.~(\ref{62}) can be
derived as follows:
\begin{eqnarray}   
 \lambda_{1,2}(\tau) &=& \sqrt{ (\Delta_s^0)^2 - 8 |K_F|^2 \pm 4|K_F|
  \sqrt{ 4|K_F|^2 + |K_s^0|^2 } } , \nonumber \\
 \lambda_{3,4}(\tau) &=& - \lambda_{1,2}(\tau) .
\label{63}
\end{eqnarray}
These eigenvalues are real for $ |K_F| < \Delta_s^0/(4\sqrt{2}) $
and describe an oscillatory solution. Two real and two pure
imaginary eigenvalues are found for  $ |K_F| >
\Delta_s^0/(4\sqrt{2}) $ reflecting the presence of amplified and
attenuated components of the fields. Using the eigenvalues $
\lambda_j $, the solution to Eqs.~(\ref{62}) can be written in a
general form using operator constants $ \hat{\alpha}_j $ and $
\hat{\beta}_j $,
\begin{eqnarray}    
 \hat{A}_{s_F}(z,\tau) &=& \sum_{j=1}^{4} \hat{\alpha}_j(\tau)
  \exp[i\lambda_j(\tau) z], \nonumber \\
 \hat{A}_{s_B}(z,\tau) &=& \sum_{j=1}^{4} \hat{\beta}_j(\tau)
  \exp[i\lambda_j(\tau) z] .
\label{64}
\end{eqnarray}

Substitution of the general solution (\ref{64}) into the second
(or the fourth) equation in (\ref{62}) provides the relations
giving the coefficients $ \hat{\beta}_j $ in terms of the
coefficients $ \hat{\alpha}_j $,
\begin{equation}   
 \hat{\beta}_j = - \frac{K_s^{0*} }{ \lambda_j + \delta_s^0/2 }
  \hat{\alpha}_j , \hspace{10mm} j=1, \ldots, 4.
\label{65}
\end{equation}
Assuming real eigenvalues $ \lambda_j $ the first (or the third)
equation in (\ref{62}) is fulfilled provided that
\begin{eqnarray}  
 \hat{\alpha}_{3,4} &=& \gamma_{1,2} \hat{\alpha}_{1,2}^\dagger , \nonumber \\
  \gamma_{1,2} &=& - \frac{4iK_F (\lambda_{3,4} +
  \delta_s^0/2)}{ \lambda_{3,4}^2 - (\delta_s^0)^2/4 + |K_s^0|^2 }.
\label{66}
\end{eqnarray}

The operator constants $ \hat{\alpha}_{1} $, $ \hat{\alpha}_{2} $,
$ \hat{\alpha}_{1}^\dagger $ and $ \hat{\alpha}_{2}^\dagger $ are
finally determined from the boundary conditions that give us the
following formulas:
\begin{eqnarray} 
 \left[ \begin{array}{c} \hat{\alpha}_{1}(\tau) \cr
  \hat{\alpha}_{2}(\tau) \cr \hat{\alpha}_{1}^\dagger(\tau) \cr
  \hat{\alpha}_{2}^\dagger(\tau) \end{array} \right] &=&
   M_1^{-1}(\tau) \left[ \begin{array}{c} \hat{A}_{s_F}(0,\tau) \cr
  \hat{A}_{s_B}(L,\tau) \cr \hat{A}_{s_F}^\dagger(0,\tau) \cr
  \hat{A}_{s_B}^\dagger(L,\tau)  \end{array} \right] , \nonumber \\
 & & \hspace{-15mm} M_1(\tau) = \left[ \begin{array}{cccc}
  1 & 1 & \gamma_1 & \gamma_2 \cr
  \vartheta_1 & \vartheta_2 &  \vartheta_3\gamma_1 & \vartheta_4\gamma_2 \cr
  \gamma_1^* & \gamma_2^* & 1 & 1 \cr
  \vartheta_3^*\gamma_1^* & \vartheta_4^*\gamma_2^* & \vartheta_1^* &
  \vartheta_2^* \end{array} \right] ;
\label{67}
\end{eqnarray}
$ \vartheta_j = - K_s^{0*} / (\lambda_j + \delta_s^0/2)
\exp(i\lambda_j L) $.

Using Eqs. (\ref{64}) and (\ref{65}) the solution for the output
operators can be written as
\begin{eqnarray}    
 \hat{A}_{s_F}(L,\tau) &=& \sum_{j=1}^{4} \hat{\alpha}_j(\tau)
  \exp[i\lambda_j(\tau) L], \nonumber \\
 \hat{A}_{s_B}(0,\tau) &=& - \sum_{j=1}^{4} \frac{K_s^{0*} }{ \lambda_j(\tau) +
  \delta_s^0/2 }\hat{\alpha}_j(\tau) .
\label{68}
\end{eqnarray}

The inverse Fourier transform of the formulas in Eq.~(\ref{68})
and return to the original operators $ \hat{a}_{s_F} $ and $
\hat{a}_{s_B} $ leaves us with the input-output relations written
in the form:
\begin{eqnarray}    
 \hat{a}_{s_b}^{\rm out}(\omega_s) &=& \sum_{c=F,B}
  \left[ \int_{0}^{\infty} d\omega_s^{'}
  U_{bc}(\omega_s-\omega_s^{'}) \hat{a}_{s_c}^{\rm in}(\omega_s^{'})
  \right. \nonumber \\
 & & \hspace{-17mm} \mbox{} \left. + \int_{0}^{\infty} d\omega_s^{'}
  V_{bc}(\omega_s+\omega_s^{'}) \hat{a}_{s_c}^{{\rm in}\dagger}(\omega_s^{'})
  \right] , \hspace{3mm} b=F,B.
\label{69}
\end{eqnarray}
The matrices $ U_{bc} $ and $ V_{bc} $ in Eq.~(\ref{69}) can be
determined numerically in general.

\section{General numerical solution and the Bloch-Messiah reduction}

To investigate the model numerically we have to replace
Eqs.~(\ref{8}) for the SSF field by their discrete variants. That
is why we introduce discrete monochromatic-mode operator
amplitudes $ \hat{A}_{s_b,i} $ that scan the spectral profiles
with period $ \Delta\omega $,
\begin{eqnarray}    
 \hat{A}_{s_b,i}(z) &=& \sqrt{\Delta \omega} \;
  \hat{A}_{a_b}(z,\omega_s^0+i\Delta\omega) ,
  \nonumber \\
 & & \hspace{10mm}   b=F,B,  \hspace{2mm} i=0,\pm 1
   \pm 2, \ldots .
 \label{70}
\end{eqnarray}
The discrete mode operator amplitudes $ \hat{A}_{a_b,i} $ obey the
usual boson commutation relations instead of those written in
Eq.~(\ref{6}). These operator amplitudes can be ordered into
vectors $ \hat{\bf A}_{s_F} $ and $ \hat{\bf A}_{s_B} $. Using
these vectors the discrete form of Eqs.~(\ref{8}) can be written
as follows:
\begin{eqnarray} 
 \frac{d}{dz} \left[ \begin{array}{c} \hat{\bf A}_{s_F}(z) \cr
  \hat{\bf A}_{s_F}^\dagger(z) \cr \hat{\bf A}_{s_B}(z) \cr
  \hat{\bf A}_{s_B}^\dagger(z)  \end{array} \right] &=&
  i {\cal M}(z) \left[ \begin{array}{c} \hat{\bf A}_{s_F}(z) \cr
  \hat{\bf A}_{s_F}^\dagger(z) \cr \hat{\bf A}_{s_B}(z) \cr
  \hat{\bf A}_{s_B}^\dagger(z)  \end{array} \right]  , \nonumber \\
 & & \hspace{-30mm} {\cal M}(z) = \left[ \begin{array}{cccc}
  {\cal D}_s& -4i{\cal K}_{sF}(z) & {\cal K}_s & 0 \cr
  -4i{\cal K}_{sF}^*(z) & - {\cal D}_s & 0 & -{\cal K}_s^* \cr
  -{\cal K}_s^* & 0 & -{\cal D}_s & 4i{\cal K}_{sB}(z) \cr
  0 & {\cal K}_s & 4i{\cal K}_{sB}^*(z) & {\cal D}_s \end{array}
  \right]. \nonumber \\
 & &
\label{71}
\end{eqnarray}
Matrix elements of the matrices $ {\cal D}_s $, $ {\cal K}_s $, $
{\cal K}_{sF}(z) $, and $ {\cal K}_{sB}(z) $ are defined as
\begin{eqnarray}  
 {\cal D}_{s,jk} &=& \delta_{jk} \frac{\delta_s(\omega_s^0+k\Delta\omega)
   }{2} , \nonumber \\
 {\cal K}_{s,jk} &=& \delta_{jk} K_s(\omega_s^0+k\Delta\omega) ,
   \nonumber \\
 {\cal K}_{sF,jk}(z) &=& \Delta\omega
   K_{nl,q}(\omega_s^0+j\Delta\omega,\omega_s^0+k\Delta\omega) \nonumber \\
 & & \hspace{-15mm} \mbox{} \times  \exp(i\delta_{nl,q}z) A_{p_F}[z,\omega_p^0+(j+k)
   \Delta\omega], \nonumber \\
 {\cal K}_{sB,jk}(z) &=& \Delta\omega
    K_{nl,q}(\omega_s^0+j\Delta\omega,\omega_s^0+k\Delta\omega)  \nonumber \\
 & & \hspace{-15mm} \mbox{} \times \exp(-i\delta_{nl,q}z)
   A_{p_B}[z,\omega_p^0+(j+k)\Delta\omega],
\label{72}
\end{eqnarray}
where $ \delta_{jk} $ means the Kronecker symbol.

As the matrix $ {\cal M} $ in Eq.~(\ref{71}) depends on $ z $,
only numerical solution is possible in general. We also need to
keep quantum features in the solution and so we have to solve the
system of linear equations (\ref{71}) for initial vectors that
form a basis. In this way, we reveal the whole evolution matrix $
{\cal U} $ that maps the operator fields at $ z=0 $ to those at $
z=L $:
\begin{eqnarray} 
 \left[ \begin{array}{c} \hat{\bf A}_{s_F}(L) \cr
  \hat{\bf A}_{s_F}^\dagger(L) \cr \hat{\bf A}_{s_B}(L) \cr
  \hat{\bf A}_{s_B}^\dagger(L)  \end{array} \right] &=&
  {\cal U} \left[ \begin{array}{c} \hat{\bf A}_{s_F}(0) \cr
  \hat{\bf A}_{s_F}^\dagger(0) \cr \hat{\bf A}_{s_B}(0) \cr
  \hat{\bf A}_{s_B}^\dagger(0)  \end{array} \right]  , \nonumber \\
 & & {\cal U} = \left[ \begin{array}{cc}
  {\cal U}_{FF} & {\cal U}_{FB} \cr
  {\cal U}_{BF} & {\cal U}_{BB} \end{array} \right] .
\label{73}
\end{eqnarray}
We note that the operator amplitudes contained in vectors $
\hat{\bf A}_{s_F}(L) $ and $ \hat{\bf A}_{s_B}(L) $ obey certain
kind of commutation relations useful in the numerical solution
(for details, see \cite{PerinaJr2005}).

Partial inversion of the linear relations in Eq.~(\ref{73})
reveals the input-output relations among the operator amplitudes,
\begin{eqnarray} 
 \left[ \begin{array}{c} \hat{\bf A}_{s_F}(L) \cr
  \hat{\bf A}_{s_F}^\dagger(L) \cr \hat{\bf A}_{s_B}(0) \cr
  \hat{\bf A}_{s_B}^\dagger(0) \end{array} \right] &=&
  {\cal U}^{\rm pinv}
  \left[ \begin{array}{c} \hat{\bf A}_{s_F}(0) \cr
  \hat{\bf A}_{s_F}^\dagger(0) \cr \hat{\bf A}_{s_B}(L) \cr
  \hat{\bf A}_{s_B}^\dagger(L)  \end{array} \right]  , \nonumber \\
 & & \hspace{-20mm} {\cal U}^{\rm pinv} = \left[ \begin{array}{cc}  {\cal U}_{FF}-{\cal U}_{FB} {\cal U}^{-1}_{BB}
    {\cal U}_{BF} & {\cal U}_{FB} {\cal U}^{-1}_{BB} \cr
    -{\cal U}^{-1}_{BB}{\cal U}_{BF} & {\cal U}^{-1}_{BB}
   \end{array} \right] .
\label{74}
\end{eqnarray}
The output operator amplitudes in the vectors $ \hat{\bf
A}_{s_F}(L) $ and $ \hat{\bf A}_{s_B}(0) $ obey the boson
commutation relations provided that the input operator amplitudes
given in the vectors $ \hat{\bf A}_{s_F}(0) $ and $ \hat{\bf
A}_{s_B}(L) $ fulfil boson commutation relations. That is why it
is convenient to rewrite the relations in Eq.~(\ref{74}) into the
form of Bogoljubov transformation. Using the operators $
\hat{a}_{s_b,i}^{\rm in} $ and $ \hat{a}_{s_b,i}^{\rm out} $
defined in analogy with their continuous counterparts we have:
\begin{eqnarray}  
 \left[ \begin{array}{c} \hat{\bf a}_{s_F}^{\rm out} \cr \hat{\bf a}_{s_B}^{\rm out}
  \end{array} \right]
  = {\bf U} \left[ \begin{array}{c} \hat{\bf a}_{s_F}^{\rm in} \cr
  \hat{\bf a}_{s_B}^{\rm in} \end{array} \right] +
  {\bf V} \left[ \begin{array}{c} \hat{\bf a}_{s_F}^{{\rm in}\dagger} \cr
  \hat{\bf a}_{s_B}^{{\rm in}\dagger} \end{array} \right]  .
\label{75}
\end{eqnarray}
The vectors $ \hat{\bf a}_{s_b}^{\rm in} $ and $ \hat{\bf
a}_{s_b}^{\rm out} $ ($ b=F,B $) are composed of the operator
amplitudes $ \hat{a}_{s_b,i}^{\rm in} $ and $ \hat{a}_{s_b,i}^{\rm
out} $, respectively.

As the matrices $ {\bf U} $ and $ {\bf V } $ describe the
Bogoljubov transformation, their Bloch--Messiah reduction
\cite{Braunstein2005,Wasilewski2006,McKinstrie2009} can be found,
\begin{eqnarray}   
 {\bf U} &=& {\bf X} {\bf \Lambda}_U {\bf Y}^\dagger , \nonumber \\
 {\bf V} &=& {\bf X} {\bf \Lambda}_V {\bf Y}^T ,
\label{76}
\end{eqnarray}
where $ \dagger $ means the hermitian conjugation and $ T $ stands
for the matrix transposition. The matrices $ {\bf \Lambda}_U $ and
$ {\bf \Lambda}_V $ are diagonal and contain real nonnegative
eigenvalues of the decomposition. The matrix $ {\bf Y} $ ($ {\bf
X} $) in Eq.~(\ref{76}) contains the right (left) eigenvectors $
{\bf Y}_i $ ($ {\bf X}_i $).

The eigenvectors defined by the Bloch-Messiah reduction give
typical modes of the nonlinear interaction and represent a
discrete form of eigenmode spectral functions $ \phi_n $ found in
Eq.~(\ref{35}) in the analytical perturbation approach.

Coefficients $ B_{s,n} $ and $ C_{s,n} $ of the generalized
superposition of signal and noise \cite{Perina1990} written for an
$ n $-th eigenmode of the Bloch--Messiah reduction (\ref{76}) and
defined in Eqs.~(\ref{49}) can be expressed using the eigenvalues
of the decomposition,
\begin{eqnarray}   
 B_{s,n} &=& {\bf \Lambda}_{U,nn}^2 \left(B_{s,n{\cal A}}^{\rm in} -1\right) +
  {\bf \Lambda}_{V,nn}^2 B_{s,n{\cal A}}^{\rm in} \nonumber \\
 & & \mbox{}  + \left( {\bf
  \Lambda}_{U,nn} {\bf \Lambda}_{V,nn} C_{s,n{\cal A}}^{\rm in} + {\rm c.c.}\right) ,
  \nonumber \\
 C_{s,n} &=& {\bf \Lambda}_{U,nn}^2 C_{s,n{\cal A}}^{\rm in} +
  {\bf \Lambda}_{V,nn}^2 C_{s,n{\cal A}}^{{\rm in}*} \nonumber \\
 & & \mbox{}  + {\bf
  \Lambda}_{U,nn} {\bf \Lambda}_{V,nn}\left( 2B_{s,n{\cal A}}^{\rm in}-1
  \right).
\label{77}
\end{eqnarray}
Symbol $ {\rm c.c.} $ stands for the complex conjugated term. The
coefficients $ B_{s,n{\cal A}}^{\rm in} $ and $ C_{s,n{\cal
A}}^{\rm in} $ related to anti-normal operator ordering
characterize the incident field of eigenmode $ n $ and can be
written as
\begin{eqnarray}   
 B_{s,n{\cal A}}^{\rm in} &=& \cosh^2(r_n) + n_{n,{\rm noise}} ,
  \nonumber \\
 C_{s,n{\cal A}}^{\rm in} &=& \frac{\exp(i\theta_n)\sinh(2r_n)}{2} ,
\label{78}
\end{eqnarray}
where $ r_n $ is the squeeze parameter, $ \theta_n $ the squeeze
phase, and $ n_{n,{\rm noise}} $ the mean number of noisy photons
in eigenmode $ n $. The number $ N_{s,n} $ of photons in eigenmode
$ n $ [$ N =\langle \hat{a}^\dagger\hat{a}\rangle $] attains a
simple form:
\begin{equation}  
 N_{s,n} = |\xi_{s,n}|^2 + B_{s,n},
\label{79}
\end{equation}
where $ \xi_{s,n} $ gives the initial coherent amplitude in
eigenmode $ n $. The principal squeeze variance $ \lambda_{s,n} $
of eigenmode $ n $ is obtained by the formula analogous to that in
Eq.~(\ref{48}) above.

The importance of the Bloch-Messiah reduction in the investigation
of squeezing is emphasized by the fact that an eigenmode with the
lowest value of the principal squeeze variance $ \lambda_{s_F,n} $
represents the solution of the optimization problem for a suitable
spectral mode profile that gives the best possible amount of
squeezing (see Appendix~A).

In the experiment, either forward- or backward-propagating fields
are interesting. Their properties can be obtained if we decompose
the eigenvectors $ {\bf X}_n $ of the Bloch--Messiah reduction
into their mutually orthogonal forward- ($ {\bf X}_{F,n} $) and
backward- ($ {\bf X}_{B,n} $) propagating parts. The corresponding
principal squeeze variances $ \lambda_{s_b,n} $ and mean photon
numbers $ N_{s_b,n} $ are given by the weighted sums of the
quantities related to the eigenvectors of the original
Bloch--Messiah reduction,
\begin{eqnarray}  
 \lambda_{s_b,n} &=& \sum_{n'} c_{b,nn'} \lambda_{s,n'}, \nonumber
  \\
 N_{s_b,n} &=& \sum_{n'} c_{b,nn'} N_{s,n'}, \hspace{5mm} b=F,B .
\label{80}
\end{eqnarray}
Using the scalar product $ \cdot $ the coefficients $ c_{b,nn'} $
are given as
\begin{equation}   
 c_{b,nn'} = \frac{ | {\bf X}_{b,n}^\dagger \cdot {\bf X}_{n'}|^2}{
  {\bf X}_{b,n}^\dagger \cdot {\bf X}_{b,n} } , \hspace{5mm} b=F,B .
\label{81}
\end{equation}

The number of effectively populated modes belongs to the most
important characteristics of the SSF field. It can be obtained
from the analysis of the amplitude correlation functions $ \langle
\hat{a}_{s,n}^{{\rm out}} \hat{a}_{s,n'}^{{\rm out}} \rangle_{\rm
vac} $ giving the correlations between the amplitudes $
\hat{a}_{s,n}^{{\rm out}} $ associated with the eigenmodes of the
Bloch--Messiah reduction. For simplicity, the correlation
functions are defined for the incident vacuum state $
|\;\rangle_{\rm vac} $ in the SSF field. Suitability of these
correlation functions for the determination of the number of
effective modes originates in the fact that it describes paired
photons in the SSF field. These paired photons are, according to
the formula for momentum operator $ \hat{G} $ in Eq.~(\ref{3}),
the elementary entities characterizing the process of
second-subharmonic generation. Using the Bloch--Messiah
decomposition in Eq.~(\ref{76}) the amplitude correlation
functions can be written as:
\begin{eqnarray}  
  \langle \hat{a}_{s,n}^{{\rm out}} \hat{a}_{s,n'}^{{\rm out}}
  \rangle_{\rm vac} &=& \delta_{nn'} {\bf X}_n {\bf \Lambda}_{V,nn}
  {\bf \Lambda}_{U,nn} {\bf X}_n^{T} .
\label{82}
\end{eqnarray}
According to Eq.~(\ref{82}) the real number $ {\bf
\Lambda}_{V,nn}{\bf \Lambda}_{U,nn} $ determines the weight of the
contribution of an $ n $-th eigenmode. After proper
renormalization of these weights guaranteeing $ \sum_{n} ({\bf
\Lambda}_{V,nn} {\bf \Lambda}_{U,nn})^2 = 1 $, the number $ K $ of
effectively populated modes is given by the cooperativity
parameter,
\begin{eqnarray}   
 K &=& \frac{ \left[  \sum_{n} ({\bf \Lambda}_{V,nn} {\bf \Lambda}_{U,nn})^2
  \right]^2 }{  \sum_{n} ({\bf \Lambda}_{V,nn} {\bf
  \Lambda}_{U,nn})^4 } .
\label{83}
\end{eqnarray}

Monochromatic frequency modes play dominant role in the
experimentally determined quantities. They can be easily evaluated
using the matrices $ {\bf U} $ and $ {\bf V} $ occurring in the
solution in Eq.~(\ref{75}). For example and assuming the incident
vacuum state in the SSF field, the amplitude frequency correlation
function $ N_{s_b s_{b'},\omega}(\omega_{s,j},\omega_{s,j'}) $ is
determined along the expression
\begin{eqnarray}   
 N_{s_b s_{b'},\omega}(\omega_{s,j},\omega_{s,j'}) &=& \langle
  \hat{a}_{s_b}^{{\rm out} \dagger}(\omega_{s,j})
  \hat{a}_{s_{b'}}^{{\rm out}}(\omega_{s,j'})
  \rangle_{\rm vac} \nonumber \\
 & & \hspace{-3.3cm} = \frac{1}{\Delta\omega} \sum_{d=F,B} \sum_{k}  V_{bd,jk}^*
  V_{b'd,j'k} , \hspace{5mm} b,b'= F,B;
\label{84}
\end{eqnarray}
$ \omega_{s,j} = \omega_s^0 + j\Delta\omega $. The matrices $
V_{bd} $ introduced in Eq.~(\ref{84}) [and similarly the matrices
$ U_{bd} $ used later] are obtained from the matrix $ {\bf V} $ [$
{\bf U} $] in Eq.~(\ref{75}) by grouping its matrix elements with
respect to the propagation direction. The spectral photon-number
density in mode $ b $ is given by the quantity $ N_{s_b
s_b,\omega}^d(\omega_{s,j},\omega_{s,j}) $.

In the time domain, amplitude correlations at two instants $ \tau
$ and $ \tau' $ are characterized by the temporal amplitude
correlation function $ N_{s_b s_{b'},\tau}(\tau,\tau') $ expressed
as
\begin{eqnarray}   
 N_{s_b s_{b'},\tau}(\tau,\tau') &=& \int d\omega_s \int d\omega'_s
  \exp(i\omega_s \tau) \exp(-i\omega'_s \tau')\nonumber \\
 & & \mbox{} \times N_{s_b s_{b'},\omega}(\omega_s,\omega'_s)
  \nonumber \\
 &=& (\Delta\omega)^2 \sum_{j,j'} \exp(i\omega_{s,j} \tau)
  \exp(-i\omega_{s,j'} \tau') \nonumber \\
 & & \mbox{} \times N_{s_b s_{b'},\omega}(\omega_{s,j},\omega_{s,j'}) .
\label{85}
\end{eqnarray}
For $ \tau = \tau' $, the quantity in Eq.~(\ref{85}) gives the
flux expressed in photon numbers.

\section{Discussion of the pulsed squeezed-light generation}

We assume that the incident forward-propagating fundamental field
is given by a Gaussian ultrashort pulse with the central
wavelength $ \lambda_p^0 = 532 \times 10^{-9} $~m and pulse
duration $ \tau_p = 1 \times 10^{-13} $~s [see Eq.~(\ref{25})]
originating in the second-harmonic frequency generation from a
pulsed Nd:YAG laser. Knowing its incident power $ P_{p_F} $ and
repetition rate $ f $ ($ f = 1 \times 10^{8} $~s$ {}^{-1} $), the
incident amplitude $ \xi_p $ defined in Eq.~(\ref{25}) is obtained
by the formula
\begin{equation} 
 \xi_p = \sqrt{ \frac{P_{p_F}L\beta_p^0}{\hbar (\omega_p^0)^2 f} }.
\end{equation}

Waveguide's depth $ t $ equals $ 5 \times 10^{-7} $~m. Its width $
\Delta y $ is $ 1 \times 10^{-6} $~m. Depth $ t_l $ of the
periodic corrugation is $ 5 \times 10^{-8} $~m which guarantees
single-mode operation at the studied frequencies (for more
details, see \cite{PerinaJr2007a}). Period $ \Lambda_l $ of this
corrugation is determined by Eq.~(\ref{10}) such that the
scattered field $ a $ fulfills the resonant condition $ \Delta_a^0
= m\pi/L $ for the first transmission peak ($ m=1 $), i.e. $
\delta_a^0 = \pm \sqrt{\pi^2/L^2 + |K_a^0|^2} $. The natural
quasi-phase mismatch $ \delta_{nl,q}^{{\rm nat},0} $ is then given
in Eq.~(\ref{44}) for the fundamental field and Eq.~(\ref{47}) for
the SSF field and determines the period $ \Lambda_{nl} $ of
nonlinear periodic poling (for details, see the end of
Subsec.~IIIB). We consider two different waveguide's lengths, $ L
= 1\times 10^{-3} $~m and $ L = 1\times 10^{-2} $~m, in the
discussion. Spectral and modal properties of the SSF field are
conveniently discussed in the shorter waveguide, that provides
wider SSF spectra. On the other hand, the nonlinear interaction is
sufficiently developed in the longer waveguide which results in
useful values of the principal squeeze variance $ \lambda $.

The section is divided into two parts. In the first part spectral
modes and their structure are discussed using the simplified model
with non-dispersion propagation developed in Sec.~IV. General
discussion of the behavior of the SSF field as it arises from
numerical solution of the model of Sec.~V is contained in the
second part.

\subsection{Spectral modes and their properties}

\begin{figure}    
 {\raisebox{3.3 cm}{a)} \hspace{5mm}
 \resizebox{0.7\hsize}{!}{\includegraphics{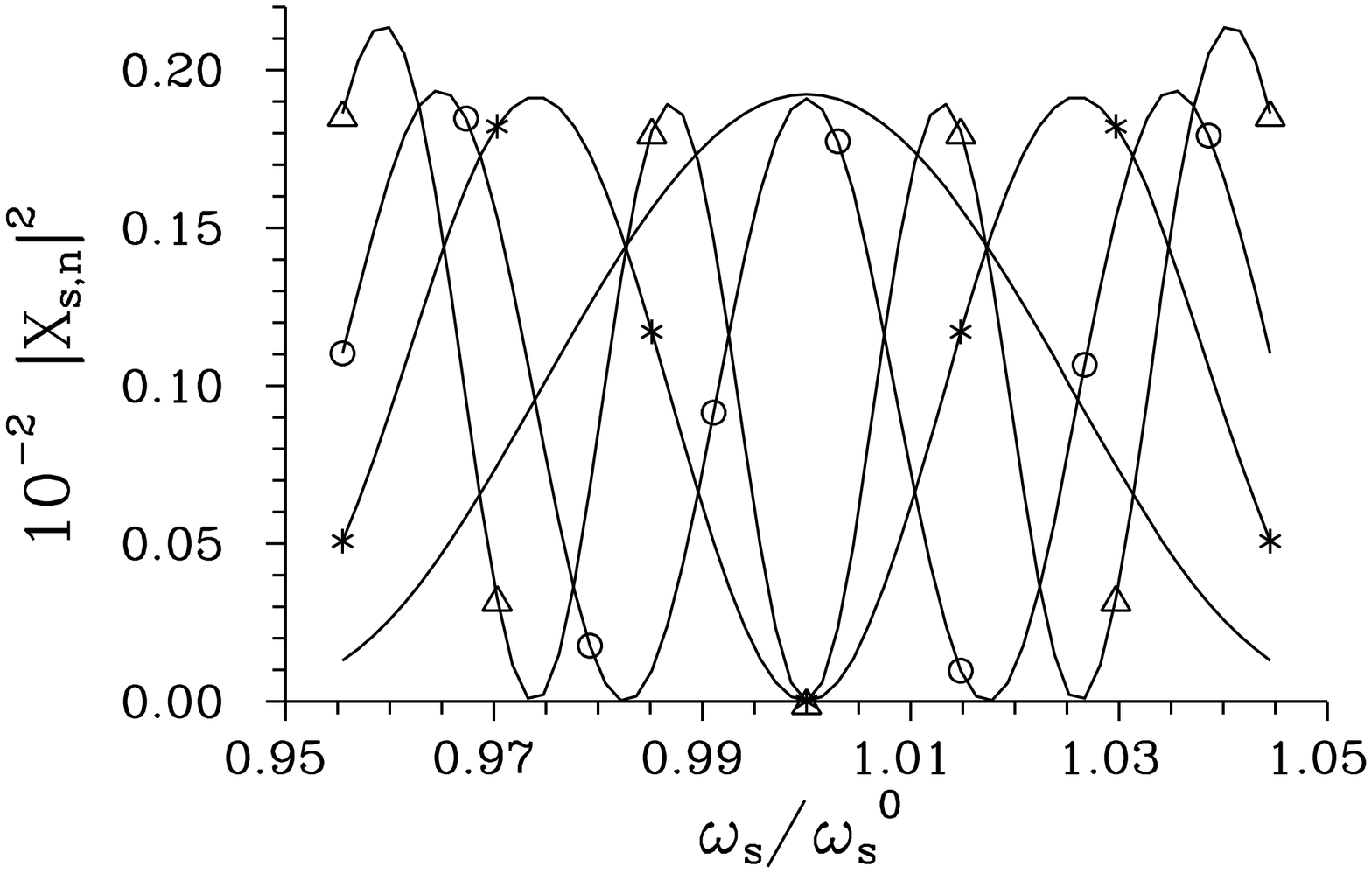}}

 \vspace{3mm}
 \raisebox{3.3 cm}{b)} \hspace{5mm}
 \resizebox{0.7\hsize}{!}{\includegraphics{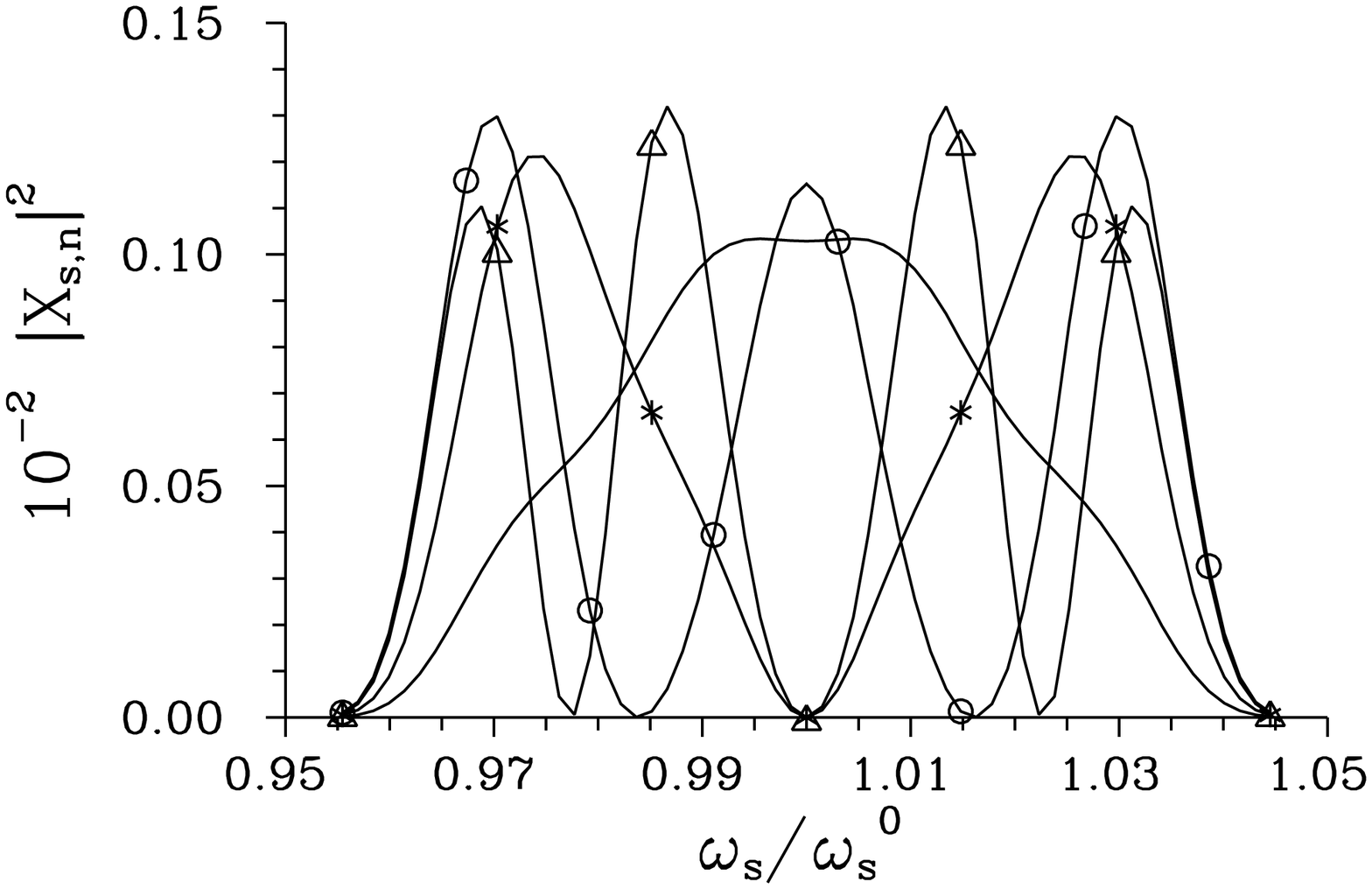}}}
 \vspace{0mm}

 \caption{Intensity spectral profiles $ |{\bf X}_{s,n}|^2 $ for the scattered
  (a) fundamental and (b) SSF field for the first (plane solid curve),
  second (solid curve with $ \ast $), third (solid curve with $ \circ $),
  and fourth (solid curve with $ \triangle $) eigenmode obtained in the model
  with non-dispersion propagation. The
  profiles are normalized such that $ \int d\omega |{\bf
  X}_{s,n}(\omega)|^2/ \omega_s^0 = 1 $. In (a) $ \Lambda_l = 1.151 \times
  10^{-7} $~m, $ \Lambda_{nl} = 3.5510 \times
  10^{-6} $~m and in (b) $ \Lambda_l = 2.459 \times
  10^{-7} $~m, $ \Lambda_{nl} = 3.5547 \times 10^{-6} $~m;
  $ P_{p_F} = 1 \times 10^{-6} $~W, $ L = 1\times 10^{-3} $~m.}
\label{fig5}
\end{figure}

The simplified model with non-dispersion propagation is useful
namely in revealing spectral properties of eigenmodes of the
nonlinear interaction. When the fundamental-field scattering is
considered, the SSF-field eigenmodes maintain qualitatively the
features obtained in the perturbation Gaussian approach in
Sec.~III resulting in the formula (\ref{35}) (see also
\cite{Bennink2002,Lvovsky2007}). Thus, an $ n $-th eigenmode ($
n=1,2,\ldots $) has $ n-1 $ zeroes in its intensity profile [see
Fig.~\ref{fig5}(a)]. Also, the larger the number $ n $ of
eigenmode, the wider the mode is.

The eigenmode structure is more complex for the scattered SSF
field that has the forward- and backward-propagating components.
Intensity profiles of these components in one eigenmode are the
same. Moreover, there exist pairs of eigenmodes with the same
intensity profile. However, they are mutually orthogonal due to
their different spectral phase profiles. Also here intensity
profiles of the components have $ n-1 $ zeros for an $ n $-th pair
of eigenmodes [see Fig.~\ref{fig5}(b)].

\begin{figure}    
 \resizebox{0.7\hsize}{!}{\includegraphics{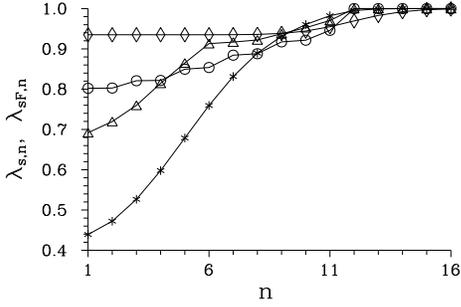}}
 \vspace{0mm}

 \caption{Principal squeeze variances $ \lambda_{s,n} $ of the maximum-squeezed
  15 eigenmodes for the scattered fundamental (solid curve with $ \ast $) and
  SSF (solid curve with $ \triangle $) field determined assuming non-dispersion
  propagation. Also the principal squeeze variances
  $ \lambda_{s_F,n} $ of the forward-propagating SSF field are shown (solid curve
  with $ \circ $). For comparison, principal squeeze variances $ \lambda_{s,n} $
  of the real waveguide without the scattered fields (solid curve with $ \diamond $)
  are drawn; $ P_{p_F} = 1 \times 10^{-6} $~W, $ L = 1\times 10^{-3} $~m.}
\label{fig6}
\end{figure}

The absence of inter-mode dispersion leads to the fact that
spectral widths of eigenmodes are given by the bandwidth of a
frequency filter used in the experiment. The fact that the
propagating monochromatic waves of the SSF field are in phase
effectively increases the nonlinear interaction. As a consequence,
smaller values of principal squeeze variances $ \lambda_{s,n} $
and greater SSF-field photon numbers $ N_{s,n} $ compared to the
real ones are predicted in the model. The principal squeeze
variances $ \lambda_{s,n} $ of the first 15 eigenmodes are drawn
in Fig.~\ref{fig6} for the scattered fundamental and SSF fields.
The principal squeeze variances $ \lambda_{s,n} $ for the
fundamental-field scattering are smaller than those for the
SSF-field scattering because the scattering in the SSF field is
weaker [compare the curves in Figs. \ref{fig2}(a) and
\ref{fig3}(a)]. In case of the scattered SSF field, even the
principal squeeze variances $ \lambda_{s_F,n} $ characterizing the
forward-propagating field and given by the formula (\ref{80}) are
shown. The values of variances $ \lambda_{s_F,n} $ can clearly be
grouped into pairs which originates in pairing of eigenmodes
discussed above. However, we note that the modes of the
forward-propagating SSF field arising from the decomposition of
eigenmodes into their forward- and backward-propagating components
are not mutually orthogonal. In fact, the number of such modes is
twice to that given by the dimension of the appropriate space. The
comparison of values of principal squeeze variances $
\lambda_{s,n} $ obtained for the waveguide with and without
scattering reveals substantial improvement caused by the
scattering (see Fig.~\ref{fig6}).

\subsection{Pulsed squeezed-light generation}

\begin{figure}    
 {\raisebox{3.3 cm}{a)} \hspace{5mm}
 \resizebox{0.7\hsize}{!}{\includegraphics{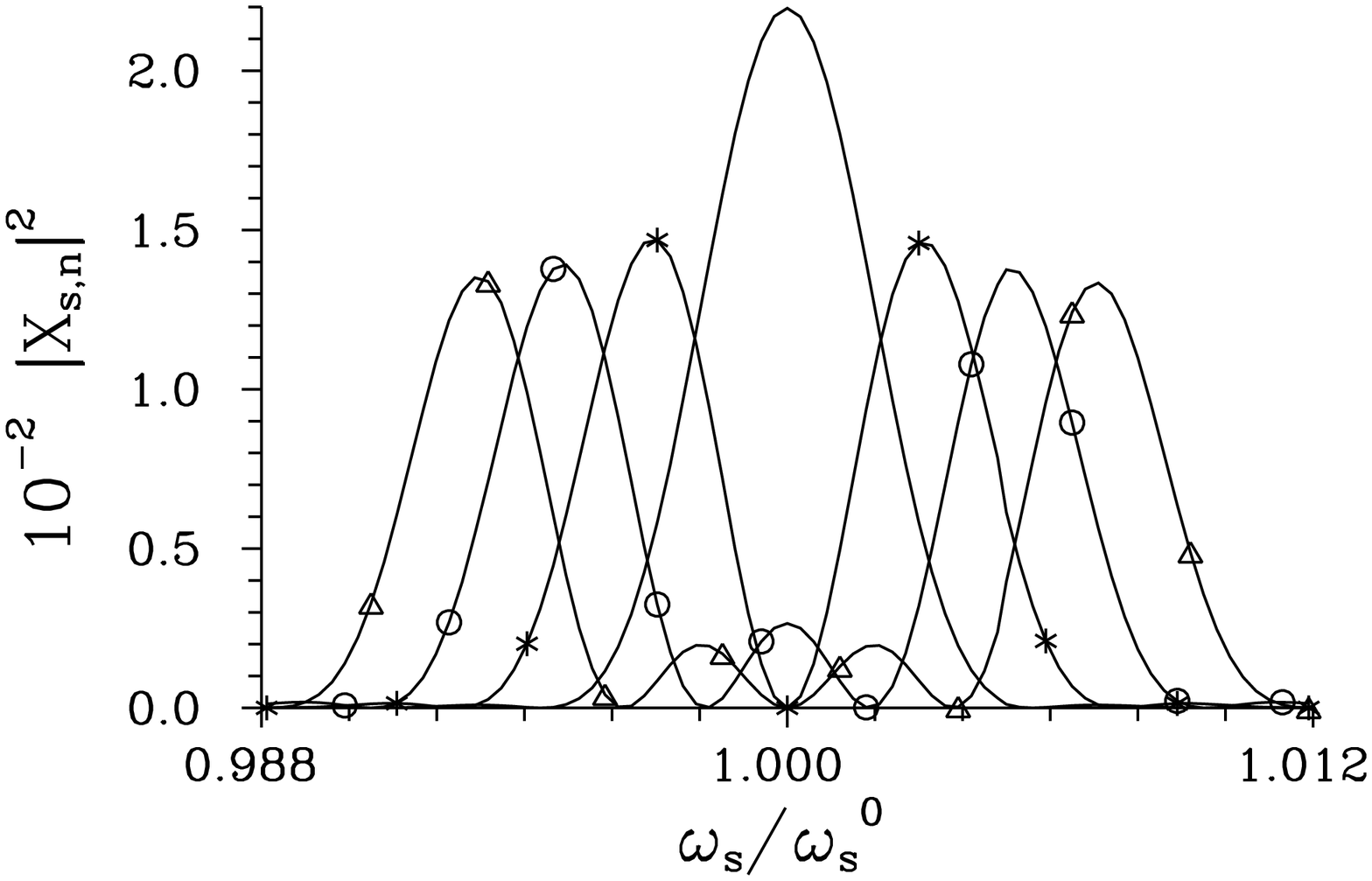}}

 \vspace{3mm}
 \raisebox{3.3 cm}{b)} \hspace{5mm}
 \resizebox{0.7\hsize}{!}{\includegraphics{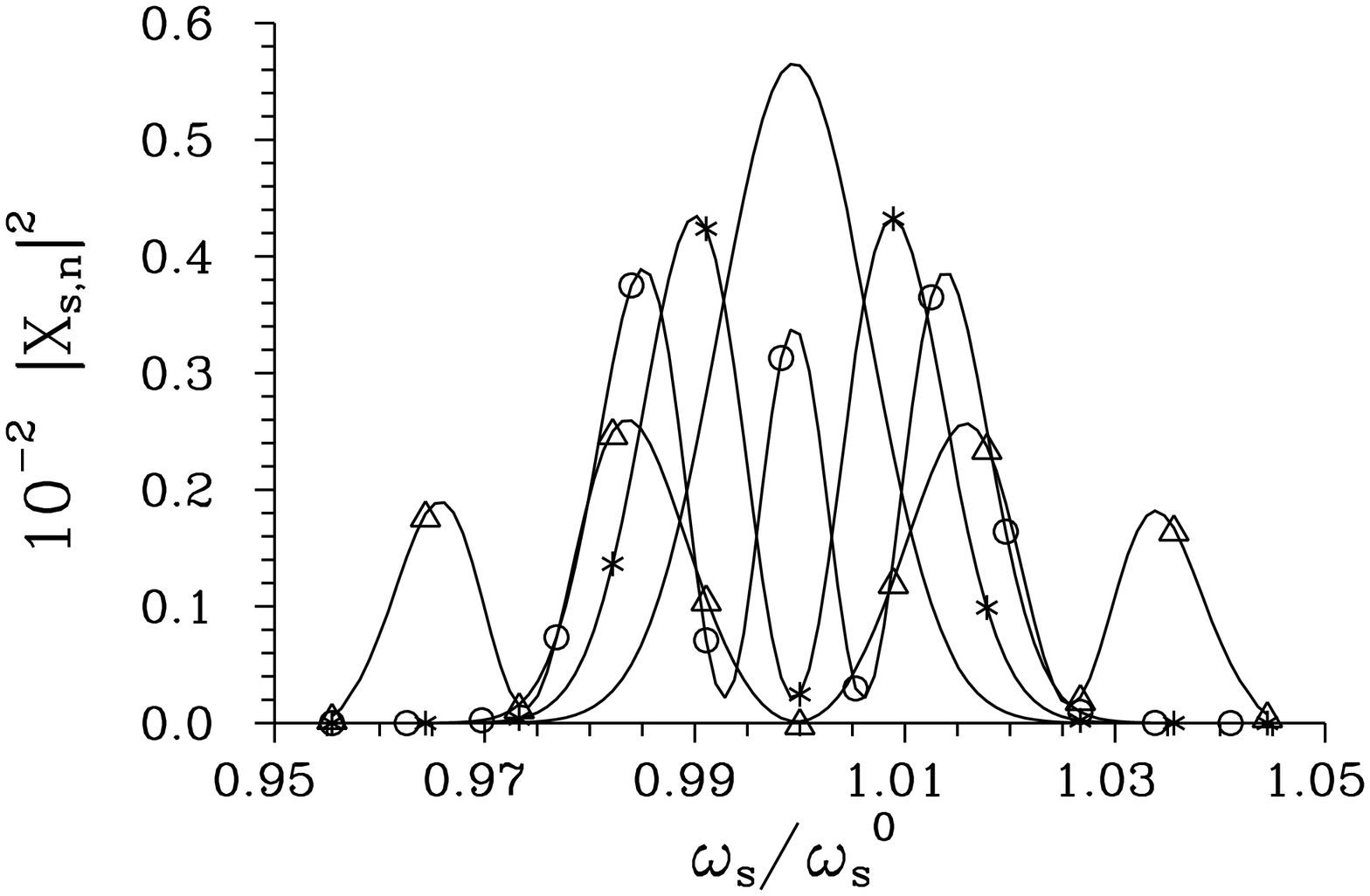}}}
 \vspace{0mm}

 \caption{Intensity spectral profiles $ |{\bf X}_{s,n}|^2 $
  for the first (solid curve without symbols),
  second (solid curve with $ \ast $), third (solid curve with $ \circ $),
  and fourth (solid curve with $ \triangle $) SSF-field eigenmode
  for (a) no scattered field and (b) scattered fundamental field.
  In (a) $ \Lambda_{nl} = 3.5516 \times 10^{-6} $~m  and
  in (b) $ \Lambda_l = 1.151 \times
  10^{-7} $~m, $ \Lambda_{nl} = 3.5510 \times
  10^{-6} $~m; $ P_{p_F} = 1 \times 10^{-6} $~W, $ L = 1\times 10^{-3} $~m.}
\label{fig7}
\end{figure}

We analyze the general solution using the model of Sec.~V assuming
the fundamental-field scattering and compare the obtained results
with those appropriate for the waveguide without scattering.
Considering the shorter waveguide, the intensity spectral profiles
of the first four eigenmodes are plotted in Fig.~\ref{fig7} for
both cases. Compared to the profiles of the model without
dispersion shown in Fig.~\ref{fig5}(a) the obtained intensity
spectral profiles are naturally bounded by spectral properties of
the waveguide (material and waveguiding dispersion, see
Fig.~\ref{fig7}). Whereas the intensity spectral profiles maintain
the appropriate number of zeros in the case without scattering,
scattering of the fundamental field leads to the replacement of
zeros by nonzero minima in these profiles. Scattering in the
fundamental field considerably broadens eigenmode spectral
profiles [compare Figs.~\ref{fig7}(a) and (b)] on one side, on the
other side it makes the overall spectra narrower [see
Fig.~\ref{fig9}(a) below]. This is caused by a complex phase
structure of the generated SSF field that requires a greater
number of eigenmodes in its decomposition. As these eigenmodes
have to be mutually orthogonal, their spectra have to be wider.
The mode structure does not significantly change when the incident
power of the fundamental field increases. This is caused by the
approximation assuming a non-depleted fundamental field.

\begin{figure}    
 {\raisebox{3.3 cm}{a)} \hspace{5mm}
 \resizebox{0.7\hsize}{!}{\includegraphics{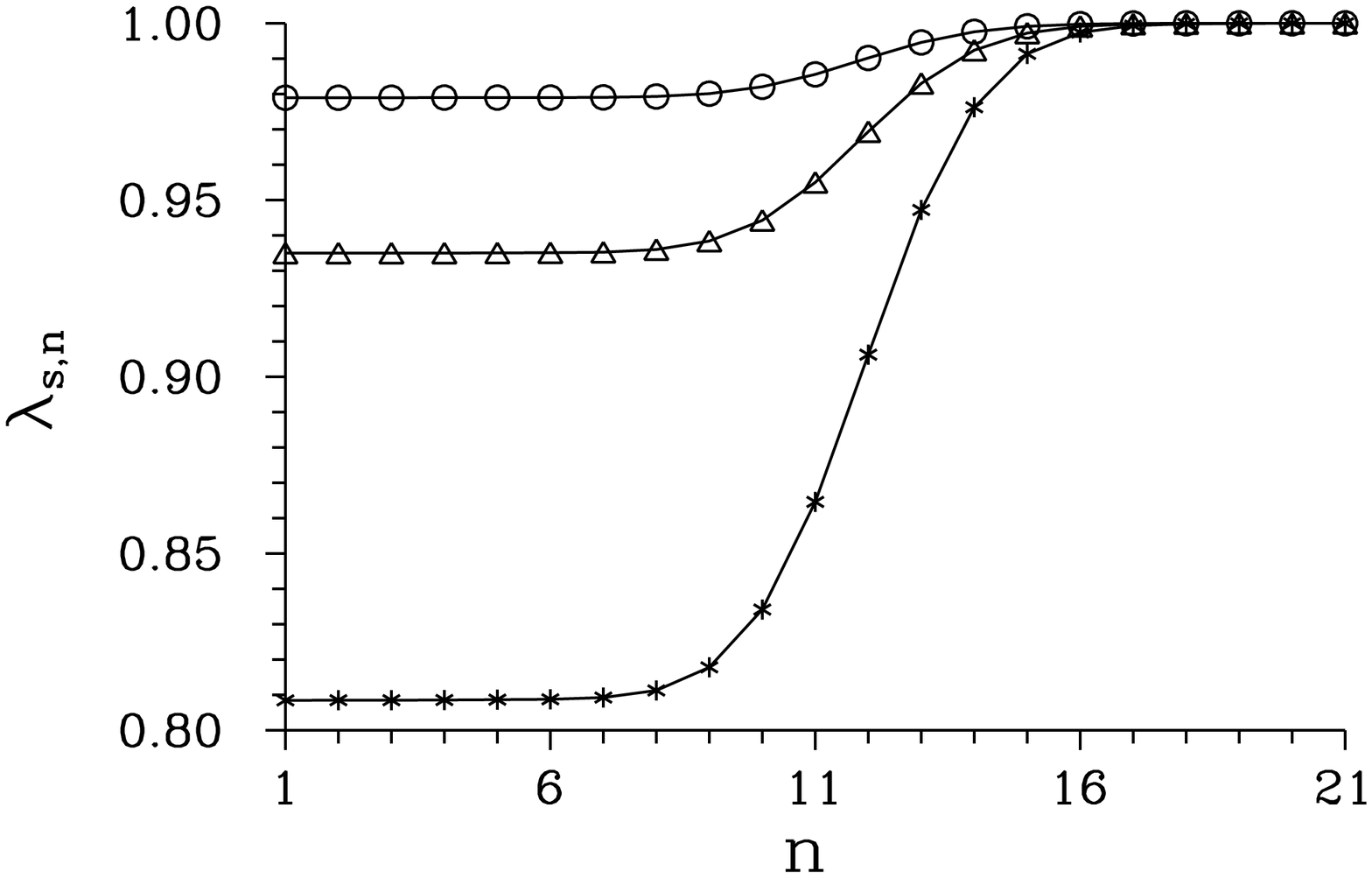}}

 \vspace{3mm}
 \raisebox{3.3 cm}{b)} \hspace{5mm}
 \resizebox{0.7\hsize}{!}{\includegraphics{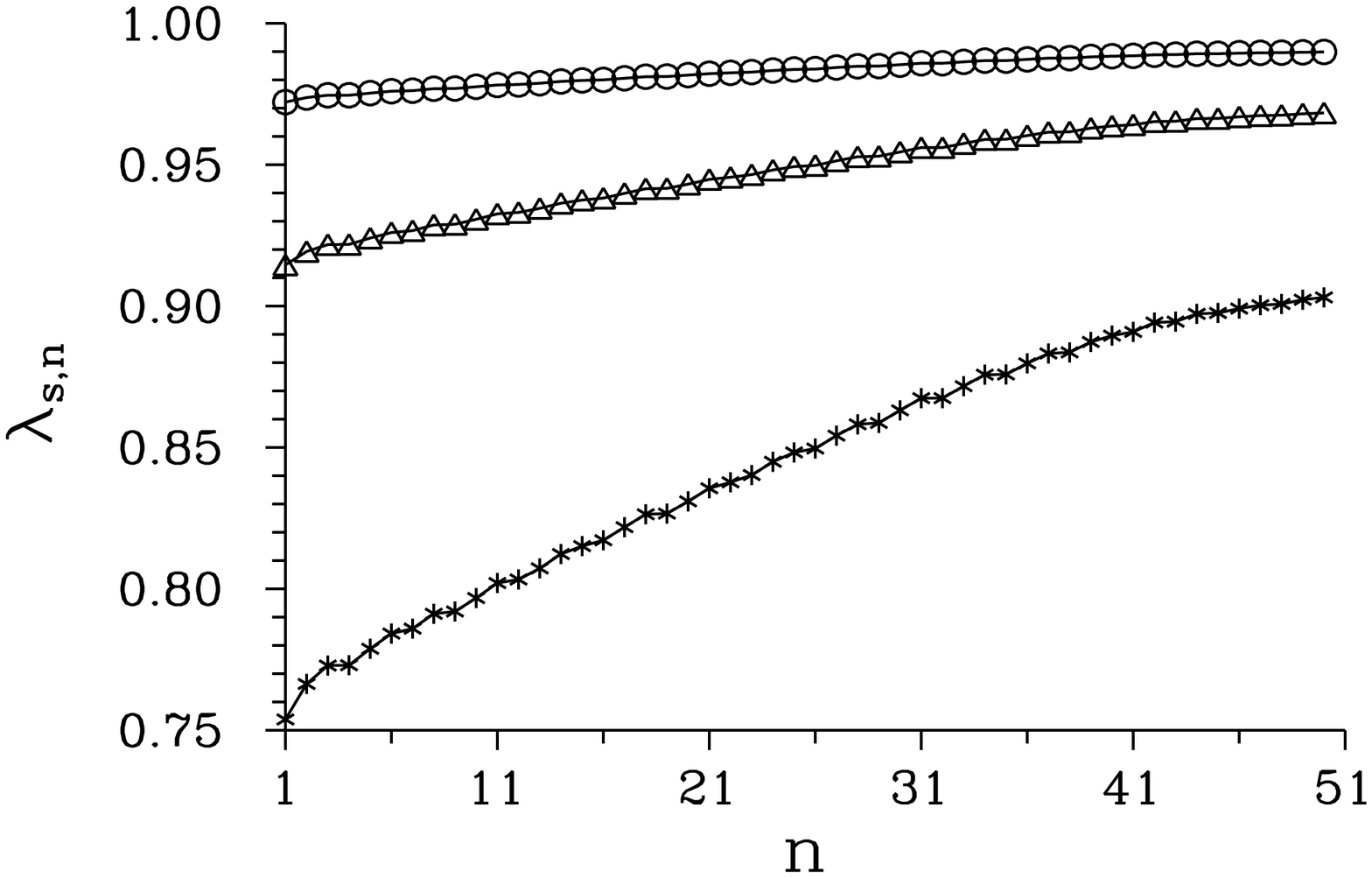}}}
 \vspace{0mm}

 \caption{Principal squeeze variances $ \lambda_{s,n} $ depending on the mode
  number $ n $ for $ P_{p_F} = 1\times 10^{-7} $~W (solid curve with $ \circ
  $), $ P_{p_F} = 1\times 10^{-6} $~W (solid curve with $ \triangle $), and $ P_{p_F}
  = 1 \times 10^{-5} $~W
  (solid curve with $ \ast $) for (a) no scattered field and
  (b) scattered fundamental field. Values of the parameters are the same as
  in the caption to Fig.~\ref{fig7}.}
\label{fig8}
\end{figure}

\begin{figure}    
 {\raisebox{3.3 cm}{a)} \hspace{5mm}
 \resizebox{0.7\hsize}{!}{\includegraphics{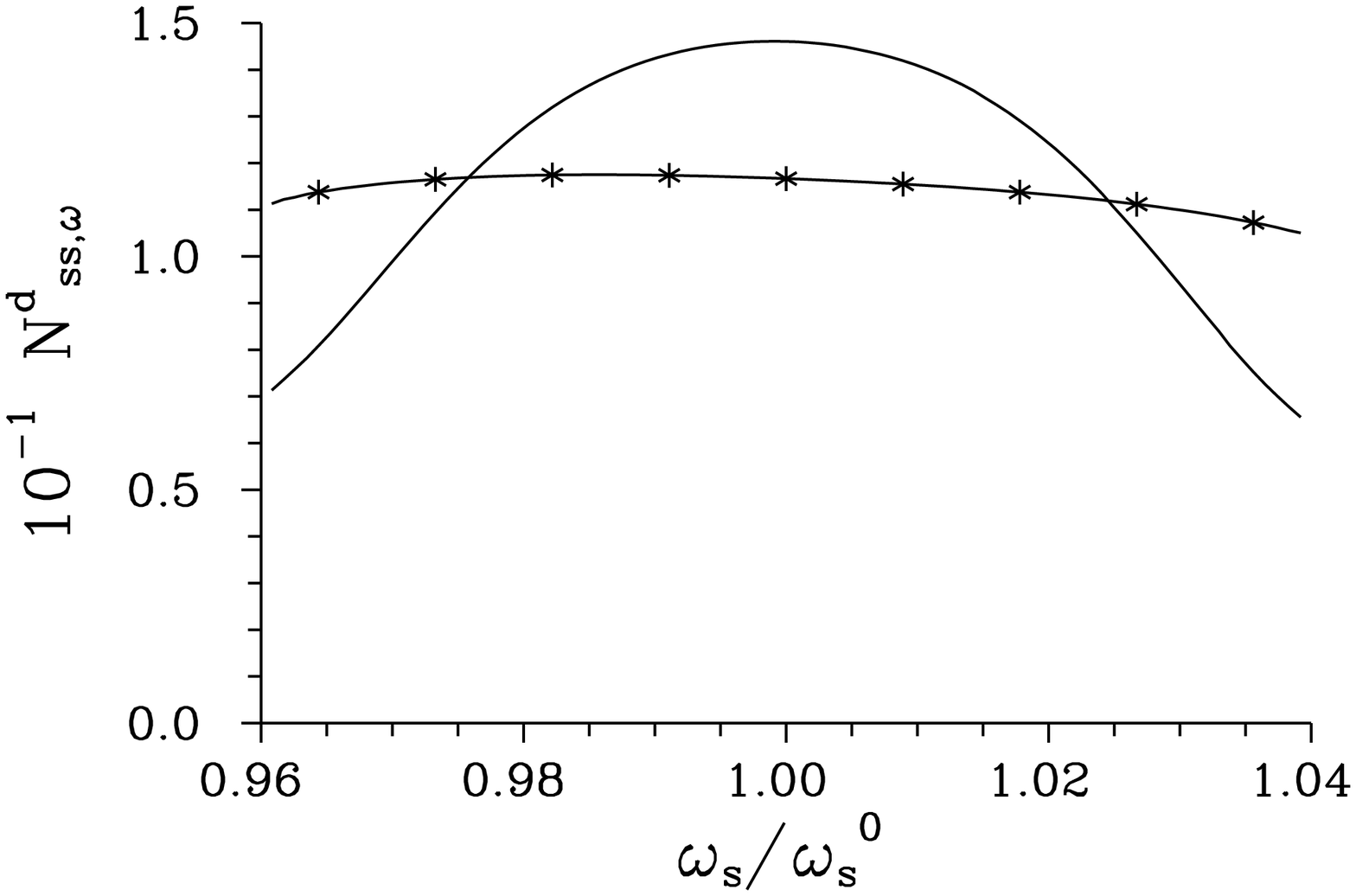}}

 \vspace{3mm}
 \raisebox{3.3 cm}{b)} \hspace{5mm}
 \resizebox{0.7\hsize}{!}{\includegraphics{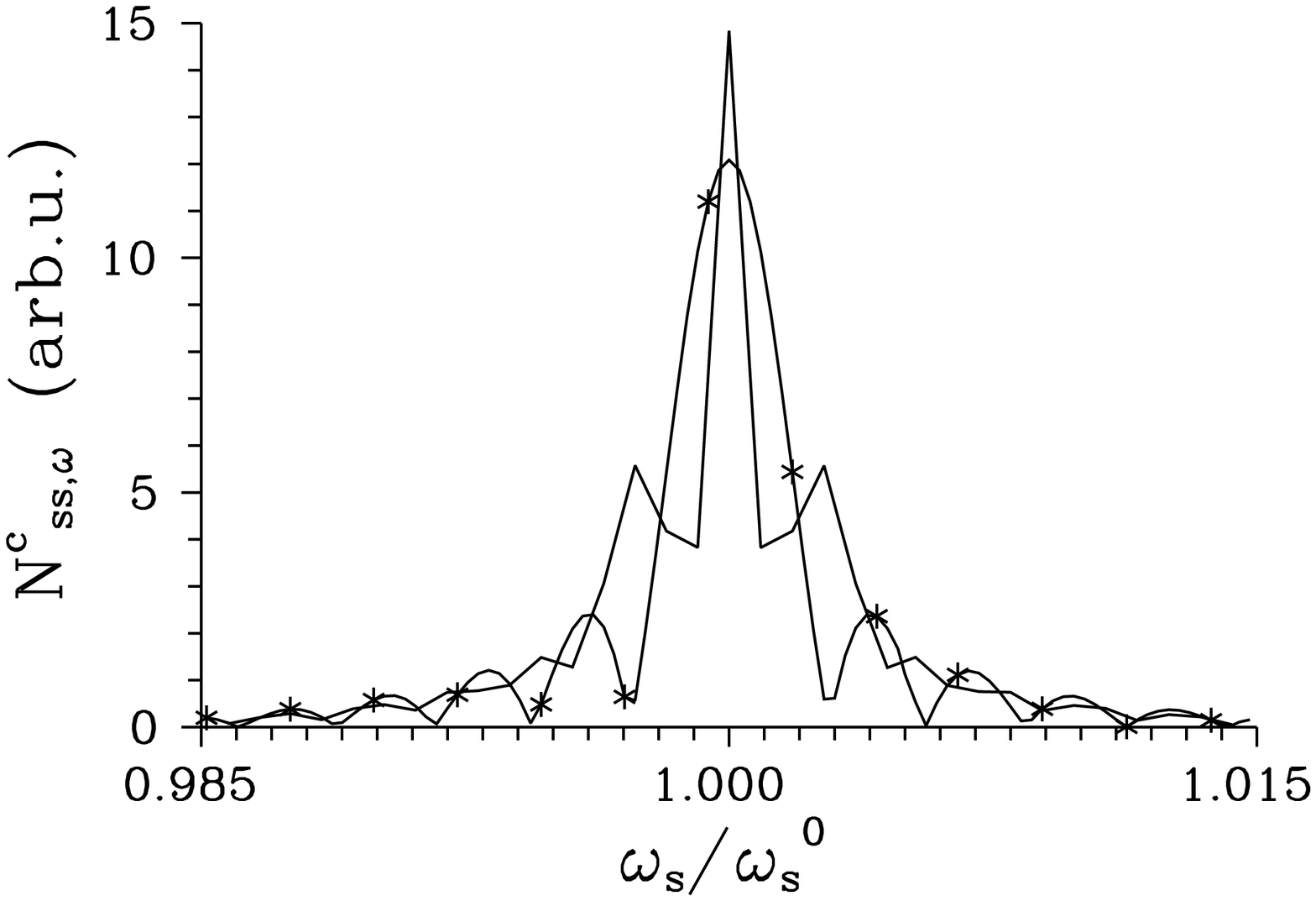}}}
 \vspace{2mm}

 \caption{(a) Intensity spectrum $ N_{ss,\omega}^d(\omega) \equiv
  N_{ss,\omega}(\omega,\omega) $ and (b) cut $ N_{ss,\omega}^c(\omega) \equiv
  |N_{ss,\omega}(\omega_s^0,\omega)| $ across the amplitude correlation function
  of the overall SSF field for the waveguide with the fundamental-field scattering
  (plane solid curve) and without scattering (solid curve with $ \ast
  $). The spectra are normalized such that $ \int d\omega
  N_{ss,\omega}^d(\omega) / \omega_s^0 = 1 $. Values of the parameters
  are the same as in the caption to Fig.~\ref{fig7}.}
\label{fig9}
\end{figure}

On the other hand, increasing values of the incident
fundamental-field power $ P_{p_F} $ decrease the principal squeeze
variances $ \lambda_{s,n} $ in all eigenmodes. This is documented
in Fig.~\ref{fig8} showing the variances $ \lambda_{s,n} $ for
three different values of the power $ P_{p_F} $. Comparison of the
curves in Figs.~\ref{fig8}(a) and (b) allows to judge the
effectiveness of scattering in the fundamental field from the
point of view of squeezed-light generation.
Scattering of the fundamental field not only increases squeezing
in the eigenmodes, it also considerably increases the number $ K $
of effectively populated eigenmodes. Whereas $ K\approx 11 $ for
the case without scattering, $ K \approx 49 $ is found for the
scattered fundamental field. Larger values of the number $ K $ of
effectively populated modes are important for pulsed homodyne
detection \cite{Bachor2004} as they lower the requirements to the
amplitude profile of the used local-oscillator field.

As the nonlinear interaction populates a larger number of
eigenmodes, the intensity spectrum $ N_{ss,\omega}^d $ of the
overall SSF field is relatively wide [see Fig.~\ref{fig9}(a)].
This is caused by nearly linear spectral dependencies of the
linear phase mismatches [see the curves in Figs.~\ref{fig2}(b) and
\ref{fig3}(b)]. On the other hand, amplitude spectral correlations
given mainly by the fundamental-field spectral width are narrow
for both considered cases [see Fig.~\ref{fig9}(b) for a cut across
the correlation function $ N_{ss,\omega}(\omega_s,\omega'_s) $].
Also spectral oscillations originating in dispersion evolution
along the $ z $ axis can be found in these correlations.

\begin{figure}    
 \resizebox{0.7\hsize}{!}{\includegraphics{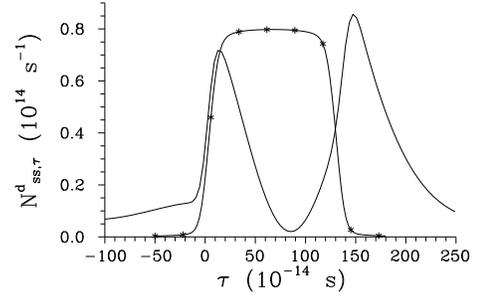}}
  \vspace{0mm}

 \caption{Flux $ N_{ss,\tau}^d(\tau) \equiv
  N_{ss,\tau}(\tau,\tau) $ of photon numbers in the SSF field
  for the waveguide with the fundamental-field scattering
  (plane solid curve) and without scattering (solid curve with $ \ast
  $). It holds that $ \int d\tau N_{ss,\tau}^d(\tau) = 1 $ and values
  of the parameters are the same as in the caption to
  Fig.~\ref{fig7}.}
\label{fig10}
\end{figure}

The SSF field is generated in the form of an ultrashort pulse (see
Fig.~\ref{fig10}). Whereas its temporal profile is close to a
rectangular shape for the case without scattering, its profile is
broken into two parts when scattering of the fundamental field is
considered. This scattering makes the SSF-field intensity spectra
narrower and, as consequence, it also extends the SSF-field
duration (from approx. 1.3~ps to 2~ps). This extension of field
duration is also caused by complex spectral phase relations
imposed by scattering of the fundamental field. Splitting of the
SSF-field pulse reflects spectral anti-symmetry around the central
frequencies $ \omega_s^0 $ and $ \omega_p^0 $. Period $ \Lambda_l
$ of linear corrugation and period $ \Lambda_{nl,1} $ of nonlinear
modulation are optimum only for the central frequencies. Whereas
their values lead to insufficient compensation on one side of the
spectrum, they overcompensate the nonlinear interaction on the
other side of the spectrum. We note that the nearly rectangular
shape in the case without scattering is given by the chosen values
of fundamental-field pulse duration and waveguide length and their
relation to the group velocities at the fundamental and SSF
central frequencies.

\begin{figure}    
 {\raisebox{3.3 cm}{a)} \hspace{5mm}
 \resizebox{0.7\hsize}{!}{\includegraphics{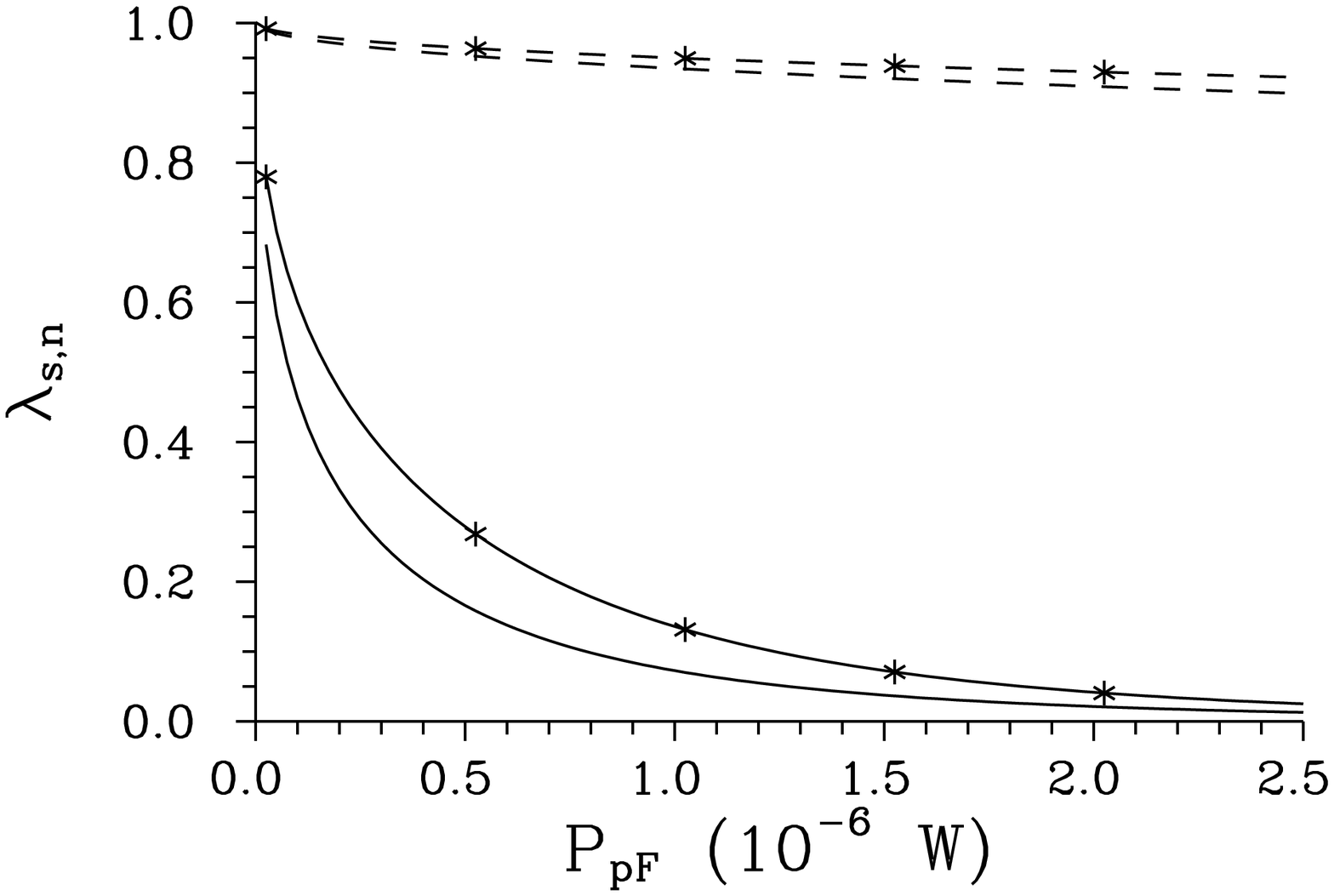}}

 \vspace{3mm}
 \raisebox{3.3 cm}{b)} \hspace{5mm}
 \resizebox{0.7\hsize}{!}{\includegraphics{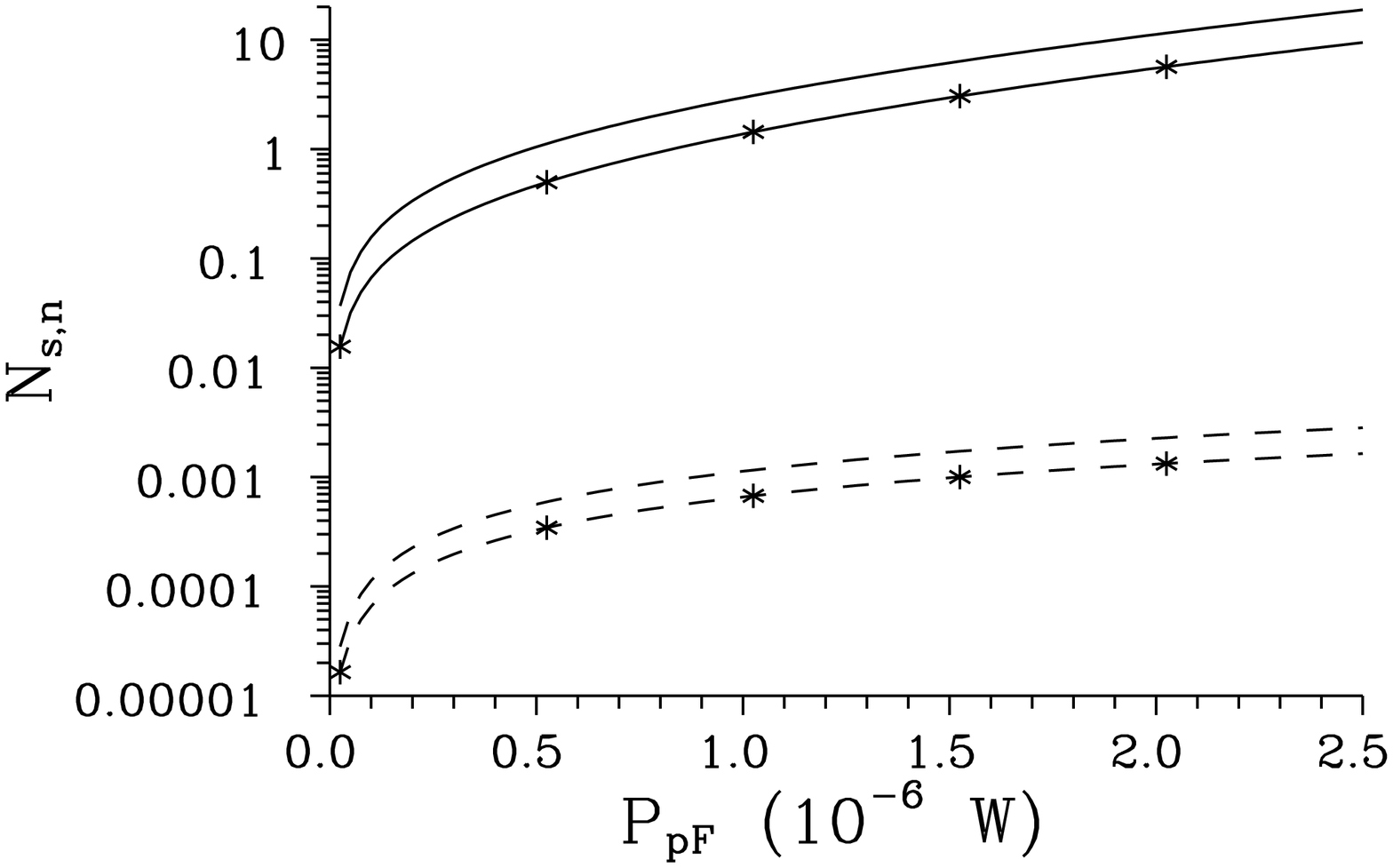}}}
 \vspace{0mm}

 \caption{(a) Principal squeeze variances $ \lambda_{s,n} $ and
  (b) numbers $ N_{s,n} $ of generated photons as they depend on incident
  fundamental-field power $ P_{p_F} $. The quantities are shown for the waveguide
  with fundamental-field scattering ($ \Lambda_l = 1.151 \times
  10^{-7} $~m, $ \Lambda_{nl} = 3.5510 \times 10^{-6} $~m, solid curves) as well as without
  scattering ($ \Lambda_{nl} = 3.5516 \times 10^{-6} $~m, dashed curves).
  Plane curves are for $ n=1 $ whereas the curves
  with $ \ast $ are for n=23 (with scattering) and n=14 (without scattering)
  equal to the number $ K $ of effectively populated modes; $ L = 1 \times
  10^{-2} $~m. In (b), the logarithmic $ y $ axis is used.}
\label{fig11}
\end{figure}

\begin{figure}    
 \resizebox{0.7\hsize}{!}{\includegraphics{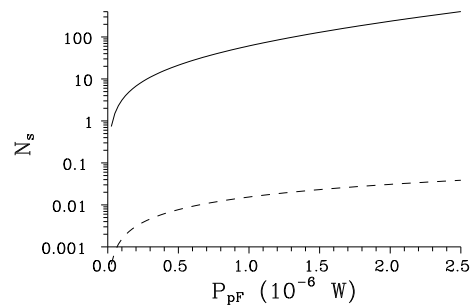}}
 \vspace{0mm}

 \caption{Numbers $ N_{s} $ of generated photons depending on incident
  fundamental-field power $ P_{p_F} $ for the waveguide
  with fundamental-field scattering (solid curve) and without
  scattering (dashed curve); $ L = 1 \times 10^{-2} $~m.
  The logarithmic $ y $ axis is used.}
\label{fig12}
\end{figure}

In order to reach useful (i.e. sufficiently small) values of the
principal squeeze variances $ \lambda_{s,n} $, either greater
fundamental-field powers or longer waveguides have to be
considered. Achievable values of the variances $ \lambda_{s,n} $
as well as numbers $ N_{s,n} $ of generated photons depend on the
fundamental-field power and are plotted in Fig.~\ref{fig11} for
the waveguide 10~mm long. As shown in Fig.~\ref{fig11}, small
values of the variances $ \lambda_{s,n} $ as well as greater
values of photon numbers $ N_{s,n} $ are practically found in all
of the first $ K $ eigenmodes.
The improvement of squeezing caused by scattering in the
fundamental field is dramatic [see Fig.~\ref{fig11}(a)].
Scattering also increases the numbers $ N_{s,n} $ of generated
SSF-field photons by more than two orders in magnitude for the
analyzed waveguide [see Fig.~\ref{fig11}(b)]. As scattering also
increases the number $ K $ of effectively populated modes (from 14
to 23) the overall number $ N_s $ of generated photons is nearly
three orders in magnitude greater when the fundamental field is
scattered (see Fig.~\ref{fig12}).

Thus, a linear periodic corrugation with suitable parameters
present in a nonlinear waveguide leads to important increase of
effective nonlinearity that results in great improvement of
amplitude squeezing of the generated light.

\section{Conclusions}

A quantum model of pulsed second-subharmonic generation in a
nonlinear waveguide with a periodic linear corrugation has been
developed. Assuming a strong fundamental field, the model has been
solved for lower second-subharmonic field intensities using
perturbation approach generalized to back-scattered fields. More
intense second-subharmonic fields with negligible inter-mode
dispersion have been treated by the Fourier-transform approach
that allows to find partly analytical solutions. Numerical
approach has been applied in the general case. Using the
Bloch--Messiah reduction spectral eigenmodes suitable for
squeezed-light generation have been revealed. Scattering by the
corrugation is more efficient in the fundamental field than in the
second-subharmonic one. Although scattering by the corrugation
makes the second-subharmonic spectra narrower, it broadens the
spectral eigenmodes. It also leads to a larger number of populated
eigenmodes. Phase relations in the nonlinear interaction imposed
by the corrugation also cause splitting of the temporal
second-subharmonic pulse. In a sufficiently long waveguide, the
corrugation dramatically increases the number of generated photons
and, hand in hand, suppresses quantum amplitude fluctuations. A
periodic corrugation thus represents a very important and
efficient tool for tailoring properties of the light generated in
modern nonlinear photonic waveguides.

\appendix

\section{An optimum mode for the pulsed squeezed light}

We look for a suitable linear combination of the output operator
amplitudes $ \hat{a}_{s,i}^{\rm out} $ that minimizes the value of
principal squeeze variance $ \lambda_{s} $. Using the
Bloch-Messiah reduction of matrices $ {\bf U} $ and $ {\bf V} $ in
Eq.~(\ref{76}) we can express this combination as follows
\cite{Luks2007x}:
\begin{equation}   
 \hat{a}_{s}^{\rm out} = \sum_{i,j} t_j {\bf X}_{ij} \hat{a}_{s,i}^{\rm out} .
\end{equation}
Coefficients $ t_j $ fulfil the normalization condition $ \sum_{j}
|t_j|^2 = 1 $. Using Eqs.~(\ref{48}) and (\ref{49}) the principal
squeeze variance $ \lambda_{s}^{\rm L} $ with the Lagrange term as
a function of $ t_j $ and $ t_j^* $ can be expressed in the
following form:
\begin{eqnarray}  
 \lambda_{s}^{\rm L}({\bf t},{\bf t}^*) &=& 1 + 2 {\bf t}^\dagger {\bf \Lambda}_V^2 {\bf t} -
  2 | {\bf t}^T{\bf \Lambda}_V {\bf \Lambda}_U {\bf t} | - \mu {\bf t}^\dagger \cdot {\bf t},
  \nonumber \\
 &=& 1 + 2 {\bf t}^\dagger {\bf \Lambda}_V^2 {\bf t} -
  2 {\bf t}^\dagger{\bf \Lambda}_V {\bf \Lambda}_U {\bf t}
  - \mu {\bf t}^\dagger \cdot {\bf t} .
\label{A2}
\end{eqnarray}
Symbol $ \mu $ in Eq.~(\ref{A2}) denotes a Lagrange multiplier
related to the normalization of vector $ {\bf t} $ composed of
coefficients $ t_j $.

Derivation of the function $ \lambda_s^{\rm L} $ in Eq.~(\ref{A2})
with respect to the coefficients $ t_i^* $ and $ t_i $ gives the
conditions:
\begin{eqnarray}   
 2 \left[ {\bf \Lambda}_V^2 - {\bf \Lambda}_V {\bf \Lambda}_U \right] {\bf t} =
  \mu {\bf t}, \nonumber \\
 2 {\bf t}^\dagger \left[ {\bf \Lambda}_V^2 - {\bf \Lambda}_V {\bf \Lambda}_U \right]  =
 \mu {\bf t}^\dagger.
 \label{A3}
\end{eqnarray}
Assuming $ \mu = 2({\bf \Lambda}_{V,ii}^2 - {\bf \Lambda}_{V,ii}
{\bf \Lambda}_{U,ii}) $ the solution to Eqs.~(\ref{A3}) is $ t_i =
1 $ and $ t_j = 0 $ for $ j \ne i $. As we look for the minimum
value of $ \lambda_s $, we choose $ i $ such that its principal
squeeze variance given by $ 1 + \mu $ is minimum.

\acknowledgments The author is obliged to M. Scalora for his
support and help. He also thanks A. Luk\v{s}, V. Pe\v{r}inov\'{a}
and O. Haderka for discussions and help with analytical
computations. This material is partly based upon the work
supported by the European Research Office of the US Army under the
Contract No. N62558-05-P-0421. J.~P. thanks the COST project
OC09026 and Operational Program Research and Development for
Innovations - European Regional Development Fund project
CZ.1.05/2.1.00/03.0058 of the Czech Ministry of Education, Youth
and Sports.

\bibliography{perina}
\bibliographystyle{apsrev}

\end{document}